%% file: main.tex
\documentclass[manuscript]{acmart}

\AtBeginDocument{%
  \providecommand\BibTeX{{%
    \normalfont B\kern-0.5em{\scshape i\kern-0.25em b}\kern-0.8em\TeX}}}

\usepackage{xcolor}
\usepackage{subcaption}
\usepackage{microtype}
\usepackage{array}
\usepackage{enumitem}
\usepackage{colortbl}
\usepackage{graphicx,calc}
\usepackage{balance}
\usepackage[utf8]{inputenc}

\newcolumntype{\$}{>{\global\let\currentrowstyle\relax}}
\newcolumntype{^}{>{\currentrowstyle}}

\def\markup{0}
\if\markup 1
\usepackage{soul}
\newcommand{\rv}[1]{{\leavevmode\color{blue}#1}}
\else
\newcommand{\rv}[1]{#1}
\newcommand{\st}[1]{}
\fi

\definecolor{symbolc}{HTML}{E60012}
\definecolor{camerac}{HTML}{28A7E1}
\definecolor{soundc}{HTML}{23AC38}
\definecolor{bangc}{HTML}{EA5413}
\definecolor{footagec}{HTML}{F8B62B}
\definecolor{endingc}{HTML}{956134}

\newcommand{\symbolandmetaphor}[1]{\textcolor{symbolc}{Symbolism and Metaphor #1}}

\newcommand{\cameraeye}[1]{\textcolor{camerac}{Camera Eye #1}}

\newcommand{\creativesound}[1]{\textcolor{soundc}{Creative Sound #1}}

\newcommand{\bigbang}[1]{\textcolor{bangc}{Big Bang #1}}

\newcommand{\oldfootage}[1]{\textcolor{footagec}{Old Footage #1}}

\newcommand{\myendingfirst}[1]{\textcolor{endingc}{Ending First #1}}

\newlength\myheight
\newlength\mydepth
\settototalheight\myheight{Xygp}
\newcommand*\inlinegraphics[1]{%
  \settototalheight\myheight{Xygp}%
  \settodepth\mydepth{Xygp}%
  \raisebox{-\mydepth}{\includegraphics[height=\myheight]{#1}}%
}

\copyrightyear{2022} 
\acmYear{2022} 
\setcopyright{acmcopyright}\acmConference[CHI '22]{CHI Conference on Human Factors in Computing Systems}{April 29-May 5, 2022}{New Orleans, LA, USA}
\acmBooktitle{CHI Conference on Human Factors in Computing Systems (CHI '22), April 29-May 5, 2022, New Orleans, LA, USA}
\acmPrice{15.00}
\acmDOI{10.1145/3491102.3501896}
\acmISBN{978-1-4503-9157-3/22/04}

\begin{document}
\title{From `Wow' to `Why': Guidelines for Creating the Opening of a Data Video with Cinematic Styles}

\author{Xian Xu}
\affiliation{%
  \institution{Computational Media and Arts Thrust,}
  \institution{The Hong Kong University of Science and Technology (Guangzhou)}
  \city{Guangzhou}
  \country{China}
}
\email{xxubq@connect.ust.hk}

\author{Leni Yang}
\affiliation{
  \institution{Department of Computer Science and Engineering,}
  \institution{The Hong Kong University of Science and Technology}
  \city{Hong Kong}
  \country{China}
}
\email{lyangbb@connect.ust.hk}

\author{David Yip}
\affiliation{%
  \institution{Computational Media and Arts Thrust,}
  \institution{The Hong Kong University of Science and Technology (Guangzhou)}
  \city{Guangzhou}
  \country{China}
}
\email{daveyip@ust.hk}

\author{Mingming Fan}
\affiliation{%
  \institution{Computational Media and Arts Thrust,}
  \institution{The Hong Kong University of Science and Technology (Guangzhou)}
  \city{Guangzhou}
  \country{China}
}
\affiliation{%
  \institution{Division of Integrative Systems and Design,}
   \institution{Department of Computer Science and Engineering}
  \institution{The Hong Kong University of Science and Technology}
  \city{Hong Kong SAR}
  \country{China}
}
\email{mingmingfan@ust.hk}
\authornote{Corresponding author}

\author{Zheng Wei}
\affiliation{%
  \institution{Business School,}
  \institution{Beijing Normal University}
  \city{Beijing}
  \country{China}
}
\email{weizheng_adams15@163.com}

\author{Huamin Qu}
\affiliation{%
  \institution{Department of Computer Science and Engineering,}
  \institution{The Hong Kong University of Science and Technology}
  \city{Hong Kong SAR}
  \country{China}
}
\email{huamin@cse.ust.hk}

\renewcommand{\shortauthors}{Xu et al.}

\begin{abstract}
Data videos are an increasingly popular storytelling form. The opening of a data video critically influences its success as the opening either attracts the audience to continue watching or bores them to abandon watching. However, little is known about how to create an attractive opening. We draw inspiration from the openings of famous films to facilitate designing data video openings. First, by analyzing over 200 films from several sources, we derived six primary cinematic opening styles adaptable to data videos. Then, we consulted eight experts from the film industry to formulate 28 guidelines. To validate the usability and effectiveness of the guidelines, we asked participants to create data video openings with and without the guidelines, which were then evaluated by experts and the general public. 
Results showed that the openings designed with the guidelines were perceived to be more attractive, and the guidelines were praised for clarity and inspiration.

\end{abstract}



\maketitle

\input{sections/1-introduction}

\input{sections/2-related-work}
\input{sections/3-style}
\input{sections/4-guideline}

\input{sections/5-evaluation}
\input{sections/6-results}
\input{sections/7-discussion}
\input{sections/8-limitations-futurework}
\input{sections/9-conclusion}


\begin{acks}
This research was supported in part by Hong Kong Theme-based Research Scheme T41-709/17N.

\end{acks}

\balance
\bibliographystyle{ACM-Reference-Format}
\bibliography{main.bib}
\end{document}

%% file: sections/1-introduction.tex
\section{Introduction}
Data videos combine diverse visual and auditory stimuli, such as motion graphics, camera movements, music, and voice-over narration, to convey data-driven insights and have been frequently employed for data storytelling~\cite{segel2010narrative,amini2015understanding}.
One critical part of a data video is its opening, which often determines whether the audience will continue watching the rest of the video or not. Thus, creating an attractive opening has been the focus of data video creators~\cite{amini2015understanding}.
The art of an opening is the art of implying and imagining through the interplay of different audio-visual elements to capture the audience's curiosity and create room for imagination.
As the master of film opening sequence Saul Bass summarized~\cite{Saul}: \textit{``My initial thoughts about what a title (opening sequence) can do was to set mood and the prime underlying core of the film’s story, to express the story in some metaphorical way. I saw the title as a way of conditioning the audience, so that when the film actually began, viewers would already have an emotional resonance with it.''}

Despite the importance of openings, little is known about how to create an attractive opening for a data video. 
Previous data video research primarily focused on the design of narrative structures~\cite{amini2015understanding, pyramid2021}, animations, and transition techniques~\cite{shi2021communicating,tang2020design,tang2020narrative}. To fill this gap, we draw inspirations from visual expressions and styles from great opening sequences of films and derive design guidance for creating attractive data video openings. 
\rv{To inform our research, we refer to previous research in data video structures~\cite{amini2015understanding, pyramid2021} and define the opening of a data video as \textit{the sequences that use visual and auditory expressions to set the tone and context of the story at the beginning of data videos.}}
\begin{figure*}
  \centering
  \includegraphics[width=\textwidth]{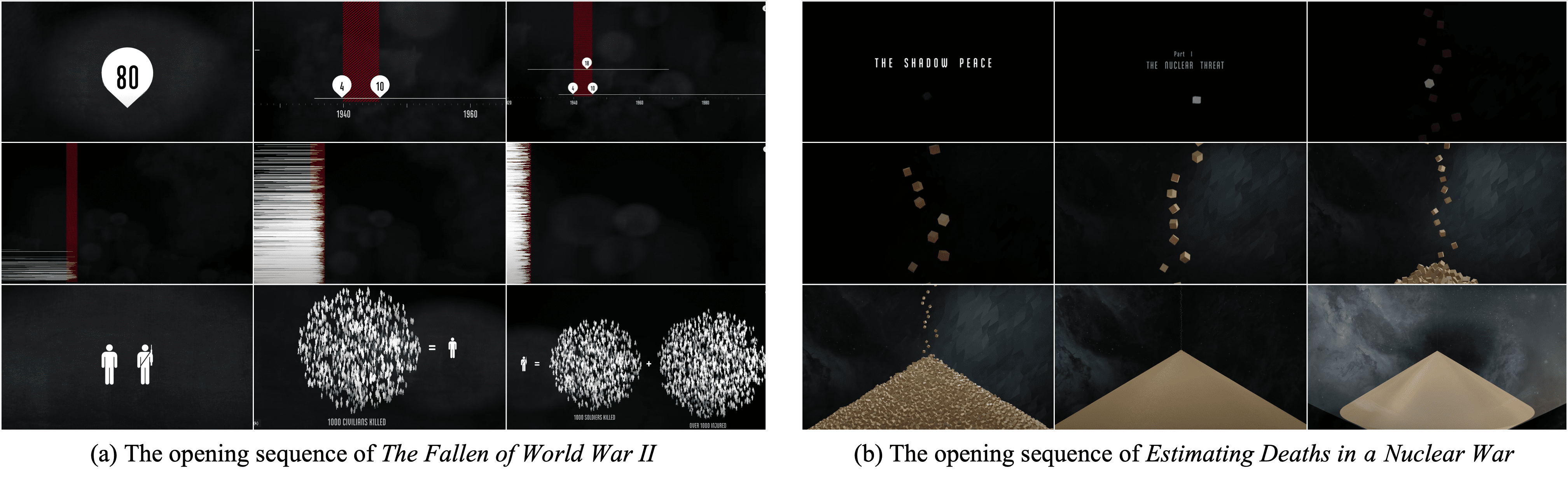}
  \caption{Opening sequences of two data videos: (a) \textit{The Fallen of World War II}~\cite{Neil2016Fallen} and (b) \textit{Estimating Deaths in a Nuclear War}~\cite{Neil2017Estimating}.}
  \label{fig:open-example}
  \Description{There are the opening screen shots of two videos. The Fallen of World War II and Estimating Deaths in A Nuclear War, which were sequentially produced by Neil Halloran, a reputable data video maker. His first video used the show-and-tell approach to introduce the audience that the video was about the number of deaths caused by the Second World War through drawing illustrative numbers and graphics on a blackboard, as shown in Fig1(a) on the left. His second video had a more cinematic style. The opening started with flashing cubes falling from off-screen, as shown in Fig1(b) on the right. The video gave a close-up shot at the falling cubes without voice-over narration but a countdown sound effect. Then it used an elegant camera movement pulling out to reveal the peak of a giant pyramid made up of these falling cubes. After that, a narration explained that those cubes represented the world population and went into the main story.}
\end{figure*}

Data videos often employ a show-and-tell approach to present an opening. Specifically, they use voice-over narration starting right from the beginning to explain the story background to the audience and use illustrative cartoons, images, and graphics to visualize the narration. Although this style efficiently sends messages, it lacks a visual style~\cite{yip2020visual} or sensational appeal that makes content richer in emotions.
On the other hand, some videos apply cinematic approaches to make a compelling opening.
One example is the evolution of styles of two videos, \textit{The Fallen of World War II}~\cite{Neil2016Fallen} and \textit{Estimating Deaths in a Nuclear War}~\cite{Neil2017Estimating}, which are sequentially produced by the \st{reputable}\rv{well-known} data video maker Neil Halloran.
His first video used the show-and-tell approach, specifically through drawing illustrative numbers and graphics on a blackboard, to introduce that the video was about the number of deaths caused by the Second World War, as shown in~\autoref{fig:open-example} (a). 
His second video has a more cinematic style. The opening started with flashing cubes falling from off-screen, as shown in~\autoref{fig:open-example} (b). The video gave a close-up shot of the falling cubes with a countdown sound effect instead of voice-over narration. Then, an elegant camera movement was used,  pulling out to reveal the peak of a giant pyramid made up of the fallen cubes. A subsequent narration explained that those cubes represented the world population and then proceeded into the main story.
The scenes of cubes created mystery among the audience, and the revelation of the giant pyramid brought a moment of awe to let the audience expect the following story.
This style of manipulating the camera movement and sound effects to create suspense and surprise is common in cinematography, and one representative work is the famous movie \textit{2001: A Space Odyssey (1969)}~\cite{2001}, as shown in~\autoref{fig:space}.
This example demonstrates how cinematic styles can add artistic value to data stories and create lasting impressions and curiosity among the audience.

Although cinematic techniques can be beneficial to improve data video openings, applying them to data videos remains challenging as no guidelines are available for data video creators to follow. Therefore, examining the styles of great cinematic openings of films and identifying ones that are adaptable to data videos are necessary so as to derive guidelines. 
Our work attained this goal with three exploratory studies. First, we, a team of a filmmaker, a screenwriter, and researchers in HCI and visualization, analyzed over 200 films from several sources such as the 100 best films from American Film Institute~\cite{afi100greatest} and \st{concluded}\rv{identified} six dominant cinematic opening styles that are adaptable to data videos. Second, we consulted experts from the film industry to further create 28 detailed guidelines for adapting the six cinematic opening styles to data videos. Finally, we conducted a comparative study in which two groups of 10 participants designed data video openings with and without the guidelines, respectively. Their designs were rated by three experts in the film industry and 20 general audiences. The results show that participants could easily follow the guidelines to create openings, and the openings created with the guidelines were rated as being more attractive. Our qualitative feedback provided further insights to understand the usability and effects of the guidelines. In summary, we make the following contributions in this work: 
\begin{itemize}[leftmargin=*]
\item We present six cinematic opening styles adaptable to data videos and 28 guidelines to realize these styles based on the analysis of the openings of best films and consultations with film experts.
\item We show the effectiveness and usefulness of the guidelines for creating attractive openings via a comparative study.
\end{itemize}

%% file: sections/2-related-work.tex
\section{Background and Related Work}
Our work is inspired and informed by prior work on \textit{narrative visualization and data videos} and \textit{cinematic openings}.
\subsection{Narrative Visualization and Data Videos}
Narrative visualization integrates storytelling techniques, data visualizations, and other visual elements to convey data-driven insights engagingly. Research in this area mainly focused on developing tools (e.g., ~\cite{amini2016authoring,kim2019datatoon,shi2020calliope}) and guidance on designing narrative visualization.
Segel and Heer~\cite{segel2010narrative} first summarized the seven basic genres of narrative visualization (i.e., \textit{Magazine style}, \textit{Annotated chart}, \textit{Partitioned poster}, \textit{Flow chart}, \textit{Comic strip}, \textit{Slideshow}, and \textit{Film/Video/Animation}) and provided a broad view of graphical and interaction techniques for narrative visualization. 
Following their work, more research studied the design factors of specific genres or data visualization types, such as data comics~\cite{bach2018design}, timeline visualization~\cite{brehmer2016timelines}, \rv{anthropographics~\cite{showingdataaboutpeople2020,showingpeoplebehinddata}}, and data-GIFs~\cite{shu2020makes}.

As a popular genre of narrative visualization, data videos have been extensively studied.
Research on data videos primarily focused on animation techniques.
Tang et al.~\cite{tang2020narrative} summarized the animated transition designs of 284 data videos to smooth the context switch to create fluent narratives.
By analyzing 70 data videos, Amini et al.~\cite{amini2016authoring} concluded eight common animation types and found that the setup animation that gradually showed the components (e.g., x-axis, y-axis, visual marks) of a visualization could increase viewers' engagement~\cite{amini2018hooked}.
Recently, by analyzing 82 data videos, Shi et al.~\cite{shi2021communicating} proposed a taxonomy of 43 animation techniques that serve for different visual narrative strategies (i.e., \textit{Emphasis}, \textit{Suspense}, \textit{Comparison}, \textit{Cohering}, \textit{Ellipsis},
\textit{Focalization}, \textit{Concretization}, and \textit{Twist}).
In addition to deepening our knowledge of animation designs in data videos, previous work provided higher-level design guidelines.
For example, Tang et al.~\cite{tang2020design} proposed guidelines for augmenting videos by integrating data visualization into scenes without causing conflicts with original content. Cao et al.~\cite{cao2020examining} introduced a broad taxonomy of the narrative and visual approaches found in 70 data videos.
\rv{Two closely related works were from Animi et al.~\cite{amini2015understanding} and Yang et al.~\cite{pyramid2021} separately identified and defined \textit{Establisher} and \textit{Setting} that set the background of stories at the beginning of data videos. 
Specifically, Amini et al.~\cite{amini2015understanding} analyzed the narrative structures of 50 videos by decomposing them and labeling the sequences of four narrative categories (i.e., \textit{Establisher}, \textit{Initial}, \textit{Peak}, and \textit{Release}) from cinematography. They further reported two major types of content (i.e., \textit{Question and New Fact}) found in the \textit{Establisher} of data videos. Recently, Yang et al.~\cite{pyramid2021} analyzed 103 data videos and concluded six types of narrative patterns (i.e., \textit{Introducing visualizations}, \textit{Statistic hook}, \textit{Preview}, \textit{Raising a question}, \textit{Introducing backgrounds}, and \textit{Presenting concrete characters}) applied in \textit{Setting}. Additionally, they concluded strategies of selecting and organizing data facts to achieve the narrative patterns and animation techniques that could enhance the presentation of narrative patterns. 
Although the two studies deepened our knowledge on designing data video openings, they focused on understanding the narrative structure of data videos. Therefore, only a few concrete guidelines were set for creating an attractive opening. 
Moreover, they were limited to summarizing narrative strategies and design techniques in existing data videos. We extend previous work by learning from cinematic art and providing concrete guidelines for making data video openings attractive with cinematic styles.}

\st{In addition to investigating certain design factors (e.g., transition techniques, animation designs), previous work also drew inspirations from documentary to improve data videos.}
\rv{Previous work has demonstrated the possibilities and benefits of adapting cinematic techniques to data videos.} 
Recently, Bradbury and Guadagno~\cite{bradbury2020documentary} found that three documentary film features---voice-of-god narration, involving the filmmaker as an actor, and indexical evidence---applied by existing data videos were beneficial.
\st{This line of work has demonstrated the possibilities and benefits of adapting cinematic techniques to data videos.} 
\rv{However, more under-explored cinematic openings styles are potentially adaptable to data videos.}
\st{, not only for the data visualization related but also for non-data situations (e.g., old footage and creative sound).} 
\rv{In addition, cinematic styles are relatively complex, involving various elements such as framing, camera movements and sound~\cite{bordwell1993film,yip2020invisible}. Thus, concrete guidelines are demanded to help data video designers manipulate them.}
\rv{To fill this gap,} we examined the best practices from well-accepted films and \st{providing design guidance}\rv{finally formulated six cinematic openings and 28 executable guidelines} on adapting these practices to create compelling data video openings.
We aim to contribute to applying the art of cinematic storytelling to data storytelling and drive more future work.

\subsection{Cinematic Opening}
We introduce background knowledge about the openings of films and related cinema studies. Unless stated otherwise, all the films discussed in our paper are feature films. Feature films are main, full-length films (usually between 75 and 210 minutes long) that are distinguished from short films such as animated cartoons and documentaries~\cite{brewster1997theatre}. For simplicity, we refer to feature film as film in the remainder of our paper.
The opening sequence of a film includes the beginning scenes that establish the setting (e.g., location, time, preview of the plot) and tone of the story and sets the audiences' expectations~\cite{wiki2021title}. Usually, an opening sequence presents the title and opening credits (key production and cast members of the film). Thus, an opening sequence is also called a title sequence. In our paper, we use the two terms interchangeably.

Most cinema studies analyzed opening sequences of cases ranging from a single film (e.g.,~\cite{ireland2018intradomain}) to a film series (e.g., James Bond films~\cite{racioppi2014geopolitics}) to a genre of film (e.g., war films~\cite{potzsch2012framing}).
They have discussed various sound and visual techniques and styles of opening sequences, such as manipulating the rhythm and volume of the music to complement the narrative progression~\cite{ireland2018intradomain}, using visual elements with symbolic meanings to foreshadow the theme ~\cite{racioppi2014geopolitics}, and framing the opening style to let the audiences see the story through an objective, subjective, or reflexive perspective~\cite{potzsch2012framing}.
However, those techniques and styles are ad-hoc, and no guidelines generalize them to broader datasets like data videos.  
Some studies have widened their analysis to larger datasets of films and focused on certain elements of opening sequences, such as creative motion graphics~\cite{bellantoni1999type} and typography to show the title and opening credits~\cite{braha2012creative}.
Finally, some comprehensive studies~\cite{inceer2007analysis,stanitzek2009reading,matamala2011opening,allison2008title} attempt to identify and classify forms and functions of opening sequences. For example, Inceer~\cite{inceer2007analysis} categorized film opening sequences into \textit{titles superimposed on a blank screen}, \textit{titles accompanied by still images}, \textit{titles accompanied with a series of moving images}, and \textit{titles built around animation and motion graphics}. Compared with previous work, we analyzed and classified cinematic openings by considering their styles that help storytellers imagine what effects their openings could achieve. Although our guidelines referenced the cinematic techniques proposed by previous work, our guidelines were tailor-made for data videos. Specifically, our guidelines were built upon visual characteristics shared by both data videos and films in style and expression.

%% file: sections/3-style.tex
\section{Cinematic Opening Styles}
The main objective of this research is to produce a set of guidelines for data storytellers to create data video openings with cinematic styles. The first step is to explore a large corpus of films and data videos and propose dominant cinematic opening styles that are adaptable to data videos. 

\subsection{Methodology}
\rv{In this section, we first explain how we identified attractive opening styles from both data videos and films and then elaborate on these opening styles.} 

\subsubsection{Data Source Collection}

\textbf{Data videos.}
\rv{We included 96 data videos that had attractive openings from two sources.}
The first source was the lists from previous studies~\cite{amini2016authoring,segel2010narrative,shi2021communicating,pyramid2021} in narrative visualizations that collected high-quality data videos. The second source was popular video platforms such as YouTube~\cite{youtube}, Vimeo~\cite{vimeo}, Douyin~\cite{douyin}, and Tencent Video~\cite{tecent}. We used keywords such as ``data stories'', ``animated infographics'', and ``data video'' to search for popular videos on these platforms. Moreover, we searched the accounts on these platforms of \rv{well-known}\st{reputable} news agencies (e.g., VOX~\cite{Vox}, New York Times~\cite{nyt}, The Economist~\cite{economist}) that produced data videos. 
\rv{All the data videos are shown in the supplemental materials.}

\textbf{Films}. 
We collected \rv{222} films from three sources. The first source was \st{a comprehensive analysis of} the 100 best films of all time from the American Film Institute~\cite{afi100greatest}. We chose this list because it was decided by a jury of 1,500 film artists, critics, and historians and has been held as one of the highest standards in the film industry and critic community. Considering that these greatest films of all time did not necessarily have a great opening, and these films might not be representative enough when it came to creative film opening, we added the second major source, which was from several popular film websites (e.g., IMDB ranking top 250~\cite{IMDB}, Hollywood~\cite{Hollywood}, Oscars~\cite{Oscars}) and YouTube channels (e.g., YouTube Movies~\cite{YouTubeMovies}) designated to rank the greatest and most creative film openings from classical to recent popular films.
The third source was professional inputs from many in-depth analyses and discussions among our research team members with film practitioners and scholars who had more than 25 years of experience in teaching university-level filmmaking and film production. 
\rv{Some of the sources we surveyed, such as YouTube channels, used the openings of TV series as representative examples of openings with cinematic styles. We did not particularly exclude these examples from our corpus. All analyzed cases can be found in the supplemental materials.}

\rv{\textbf{Exclusion Criteria}. To identify opening styles transferable to data videos, we analyzed the openings of films and excluded \textit{character-driven openings} and \textit{dialogue-driven openings} that were unsuitable to data videos. A character-driven opening starts with a situational scene that primarily involves the thinking, emotions, actions, and reactions of a character and leverages the appearance of the character (e.g., costume, hair, makeup). 
For example, the character-driven opening of \textit{Toy Story (1995)}~\cite{toystory} introduces the boy and his toy Woody.
A dialogue-driven opening involves a dialogue among characters. For example, in the dialogue-driven opening of \textit{Before Sunset (2004)}~\cite{beforesunset}, the main character, Jesse, is having a long dialogue with other characters.}

\rv{
We did not include documentaries in our corpus because the key authoritative ranking sources we used did not include any documentaries. However, excluding documentaries unlikely affected the identification of cinematic opening styles from the film industry, because if a documentary wants to attract the audience's attention with an attractive opening, it often turns to films to draw inspiration~\cite{bordwell1993film}.}

\subsubsection{\rv{Analysis Method}}
Prior research did not provide any reference model for cinematic opening styles. As a result, two teams of the authors analyzed data videos and films to derive six cinematic opening styles that were adaptable to data videos. 
Specifically, one team (the first and the second authors) analyzed the openings of 96 data videos, and the other team (the first and the third authors) analyzed the openings of over 200 films. 



The two teams adopted thematic analysis~\cite{braun2006using} to code the opening styles of the data videos and films separately. After initial coding, the two teams met and discussed their coding results for at least three rounds to identify cinematic styles of which related approaches were applied by existing data videos. \rv{While all coders identified new opening styles at the early stage of their analyses, they stopped finding new opening styles toward the end of their analyses. This suggested that their analyses reached saturation, and the identified opening styles were reasonably complete.} Finally, they reached a consensus on six cinematic opening styles. This process ensured that the six cinematic opening styles were adaptable to data videos. \rv{In addition, in our follow-up user studies, we found that all participants, including film experts, did not identify new opening styles.} The detailed coding results can be found in the supplementary materials.


\subsection{Six Cinematic Opening Styles}
In this section, we explain the six cinematic opening styles with examples drawn from films and data videos: \textit{Symbolism and Metaphor}, \textit{Camera Eye}, \textit{Creative Sound and Music}, \textit{Big Bang}, \textit{Old Footage}, and \textit{Ending First}. It is worth noting that many examples cited below used more than one of these six styles\rv{, suggesting that these six opening styles can be used in combination.}
\rv{One may be concerned that the length difference between full-length feature films and much shorter data videos may prevent the use of cinematic opening styles in data videos. However, our analysis of the film corpus indicates that a cinematic style is independent of its length. For example, the Big Bang style can be shown in one shot with only a few seconds long as in \textit{Casino (1995)}~\cite{Casino}. It can also be shown in a shot sequence lasting more than a minute as in \textit{Apocalypse Now (1978)}~\cite{Apocalypse}. Similarly, both the opening shots of \textit{Hard Boiled (1992)}~\cite{hard} and \textit{Raging Bull}~\cite{RagingBull} belong to Symbolism and Metaphor. The opening shot of the former lasts only a few seconds while the latter lasts a few minutes. This suggests that cinematic styles are adaptable to data videos of different duration.}

\begin{figure}[!ht]
  \centering
  \begin{minipage}[b]{0.45\textwidth}
      \includegraphics[width=\textwidth]{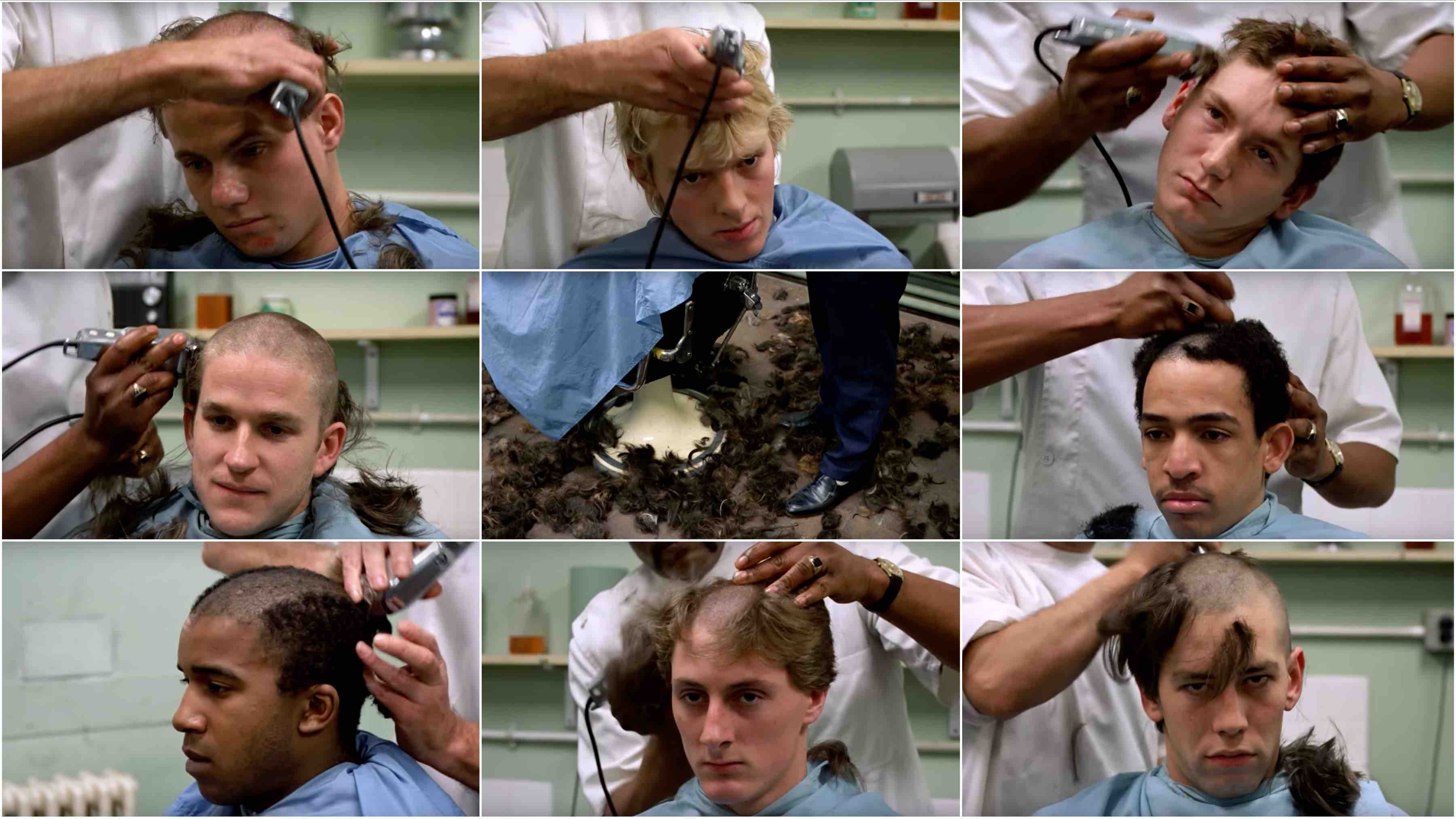}
    \caption{The opening sequence of \textit{Full Metal Jacket (1987)} using \textit{Symbolism and Metaphor}~\cite{game2013full}.}
    \label{fig:metal}
    \Description{This is the opening sequence of \textit{Full Metal Jacket (1987)} using \textit{Symbolism and Metaphor}~\cite{game2013full}. The famous haircutting opening scene in Stanley Kubrick's \textit{Full Metal Jacket (1987)}~\cite{game2013full}, in which a group of young army recruits were shaved all their hair, symbolized the stripping of their individuality.}
  \end{minipage}
  \hfill
  \begin{minipage}[b]{0.45\textwidth}
     \includegraphics[width=\textwidth]{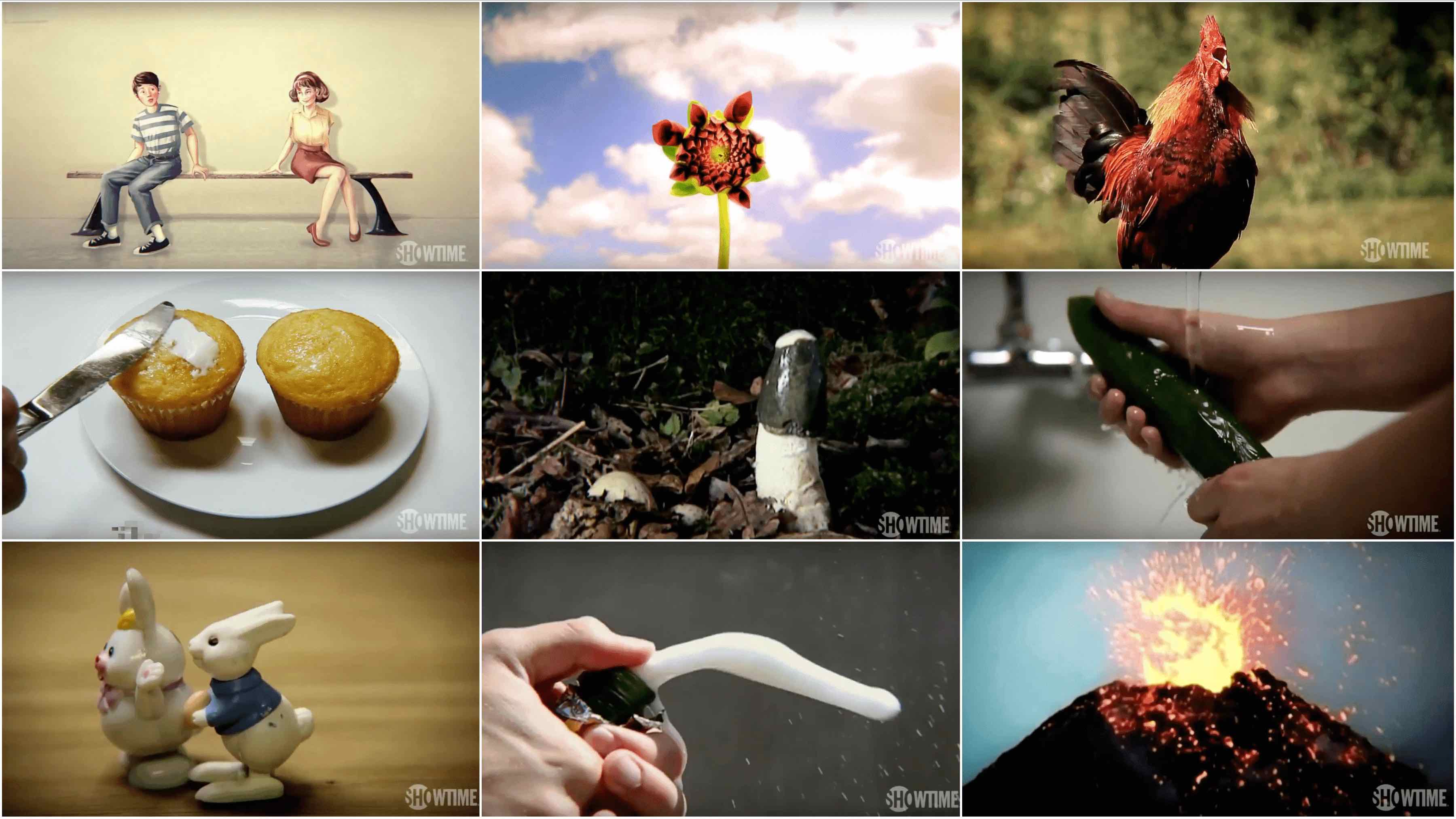}
    \caption{The opening sequence of \textit{Master of Sex (2013)} using \textit{Symbolism and Metaphor}~\cite{masters}.}
    \label{fig:master}
    \Description{This is the opening sequence of \textit{Master of Sex (2013)} using \textit{Symbolism and Metaphor}~\cite{masters}. The sequence shows a series of implicit images to subtly suggest the subject of sex.}
  \end{minipage}
\end{figure}

\begin{figure} [!ht]
  \centering
  \begin{minipage}[b]{0.45\textwidth}
  \includegraphics[width=\linewidth]{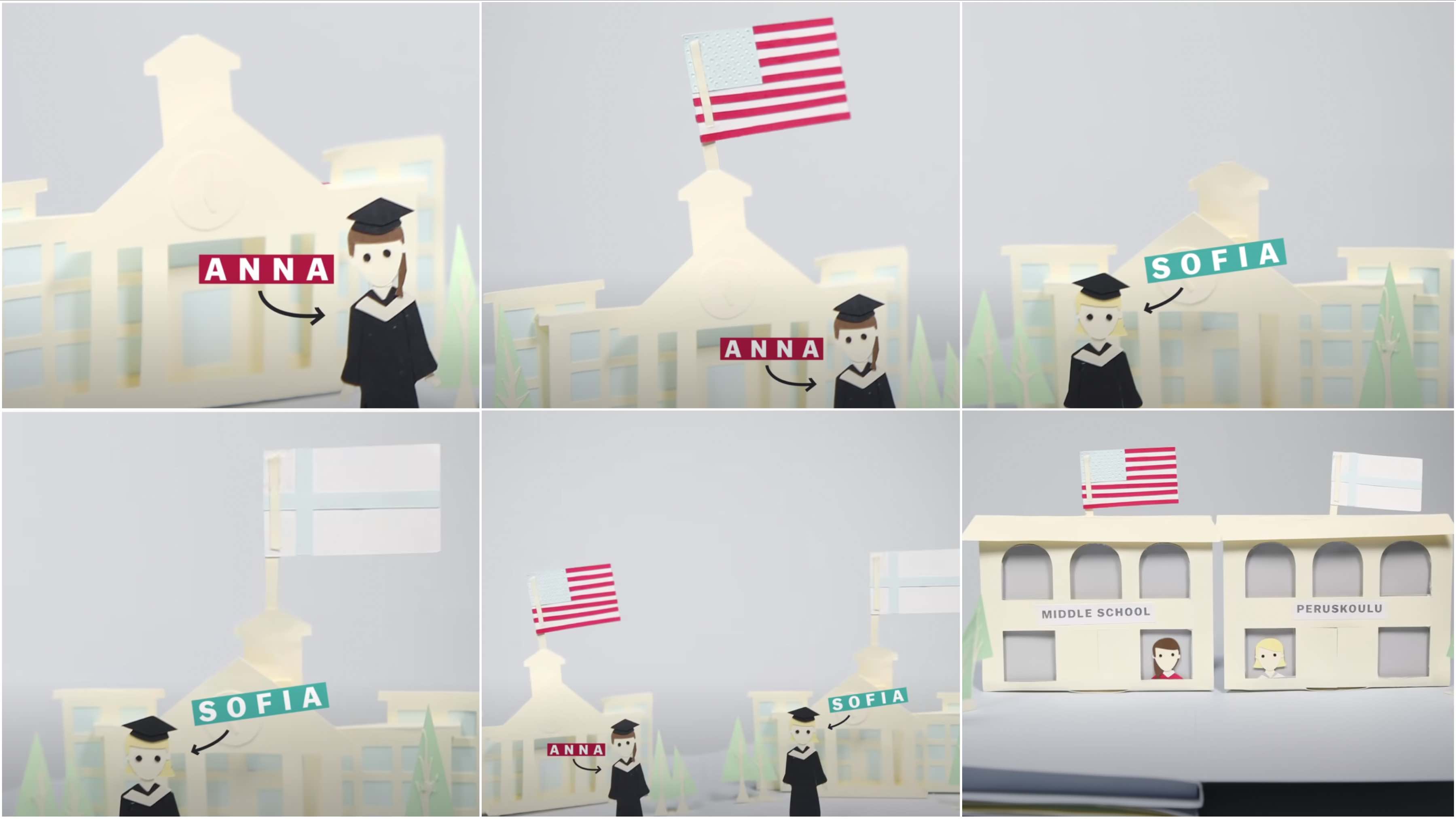}
  \caption{The opening sequence of \textit{Teaching in the US vs. the Rest of the World} using \textit{Symbolism and Metaphor}~\cite{Teaching}.}
  \label{fig:teaching}
  \Description{This is the opening sequence of \textit{Teaching in the US vs. the Rest of the World} using \textit{Symbolism and Metaphor}~\cite{Teaching}. It told a story of the teaching experiences of two imaginary figures, Anna from USA and Sophia from Finland, to symbolize the story topic.}
  \end{minipage}
  \hfill
  \begin{minipage}[b]{0.45\textwidth}
    \includegraphics[width=\linewidth]{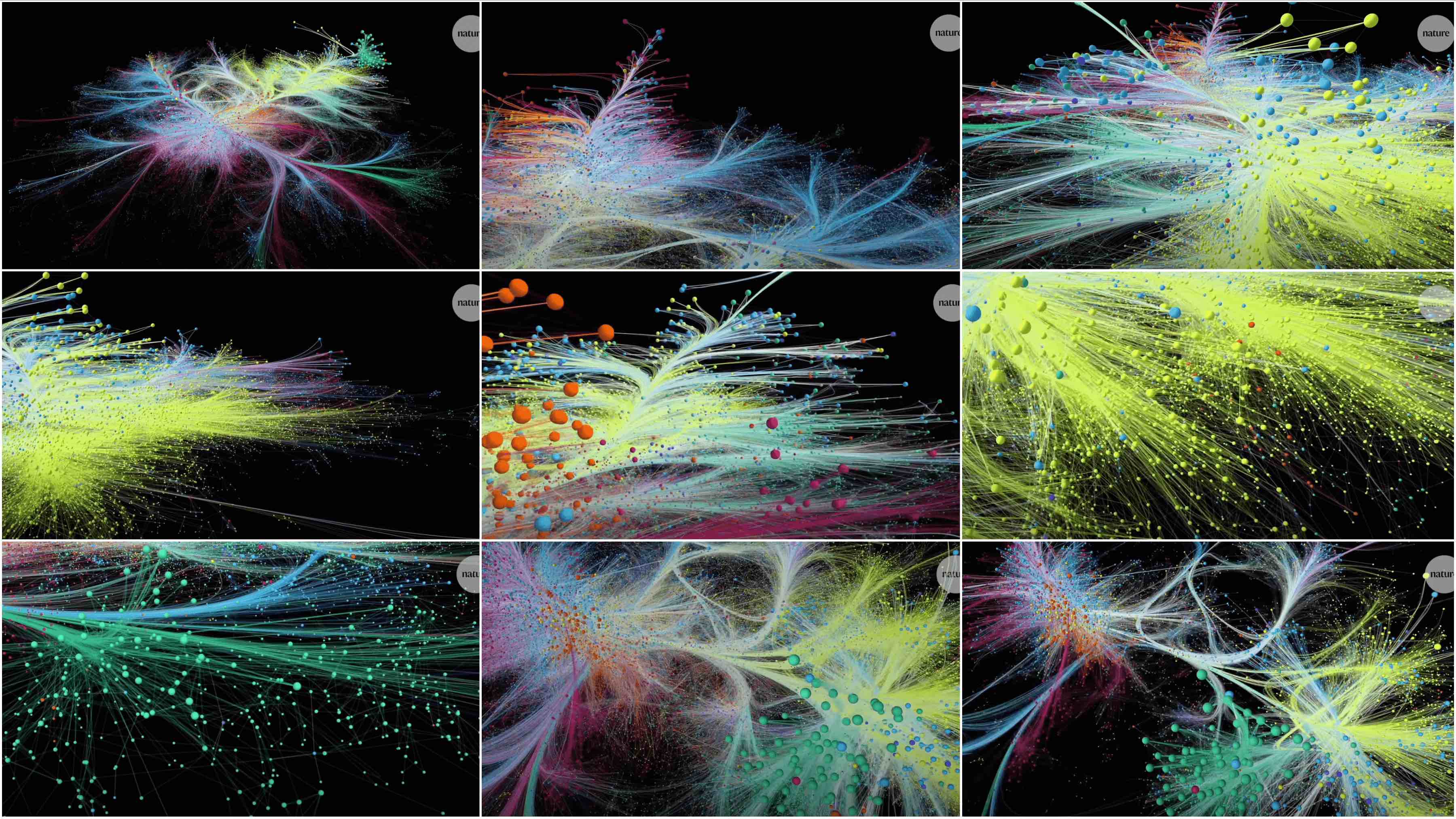}
  \caption{The opening sequence of \textit{A Network of Science: 150 Years of Nature} using \textit{Camera Eye}~\cite{network}.}
  \label{fig:network}
  \Description{This is the opening sequence of \textit{A Network of Science: 150 Years of Nature} using \textit{Camera Eye}~\cite{network} which also experimented with the \textit{Camera Eye} style to introduce a complex network by showing different parts of the network in a series of short shots.}
  \end{minipage}
\end{figure}

\textbf{\symbolandmetaphor}: This style uses concrete, familiar, and relevant objects and colors to visualize the key theme, emotion, and concept of the story to attract the audience.
For example, the famous haircutting opening scene in Stanley Kubrick's \textit{Full Metal Jacket (1987)}~\cite{game2013full}, in which a group of young army recruits was shaved all their hair, symbolized the stripping of their individuality, as shown in~\autoref{fig:metal}.
\textit{Apocalypse Now (1978)}~\cite{Apocalypse} used the explosion of the jungle to symbolize the bombing of Vietnam to foreshadow the subject of war.
A recent example is the opening sequence of \textit{Master of Sex (2013)}~\cite{masters} that showed a series of implicit images to subtly suggest the subject of sex, as shown in~\autoref{fig:master}.
Not only for films but also data videos, \textit{Symbolism and Metaphor} first involves encoding the storytellers' symbolic meaning to an image, which is then decoded by the audience as intended. Although the opening of films often involves protagonist and antagonist characters and data stories do not always involve them, data stories can use this style to tell an expressive story behind the data. For example, the opening of the data video \textit{Teaching in the US vs. the Rest of the World}~\cite{Teaching} told a story of the teaching experiences of two imaginary figures, Anna from the USA and Sophia from Finland, to symbolize the story topic, as shown in~\autoref{fig:teaching}. \rv{A similar approach has been found in anthropographics~\cite{showingdataaboutpeople2020,showingpeoplebehinddata}, a type of static visualization using human icons or figures as symbols. By contrast, the \textit{Symbolism and Metaphor} opening style uses not only people but also objects to explore symbolism's applications to data videos.}


\textbf{\cameraeye}: This style reveals the story context by using expressive camera movements, such as long take, various compositions, static and movement shots, to guide the audience's attention to specific areas or objects in the opening scenes. 
This style helps build tension and suspense to increase the audience's engagement and expectation, as many famous film directors such as Alfred Hitchcock, Orson Welles, Alfonso Cuarón, and Sam Mendes have done in their films. 
One common approach to the style is placing a \textbf{long opening shot} at a special place and angle for narrative purposes. For example, \textit{Fight Club (1999)}~\cite{fight} was about internal problems created by split personality disorder. To visualize this mental disorder, the camera literally took the audience in one long shot to the internal of the character's brain, where all his psychological problems originated. 
Another example was done by the same design team of \textit{Fight Club} in \textit{Lord of War (2005)}~\cite{Lord}, a dramatic movie about illegal gun trades. The long camera shot in the opening sequence accompanied by Springfield's song \textit{For What It's Worth (1967)}~\cite{forwhat} showed in one long take the making of a bullet from scratch to its final destiny of killing life.
Whether it is a physical camera in film or a virtual camera in data video, the principle of camera work and movement is the same. 
For example, a long uninterrupted shot help add cinematic qualities to data videos by maintaining a seamless flow and consistent style of storytelling, as demonstrated in the work of Neil Halloran that is discussed in the Introduction section, as shown in~\autoref{fig:open-example} (a).
Another case is the opening of the data video \textit{A Network of Science: 150 Years of Nature}~\cite{network}. The video experimented with the \textit{Camera Eye} style to introduce a complex network by showing different parts of the network in a series of short shots, as shown in~\autoref{fig:network}. However, it can more clearly show the composition of the network by using a long shot to show the spatial relationship between the parts and the whole visualization.

\begin{figure}[!ht]
  \centering
  \begin{minipage}[b]{0.45\textwidth}
    \includegraphics[width=\linewidth]{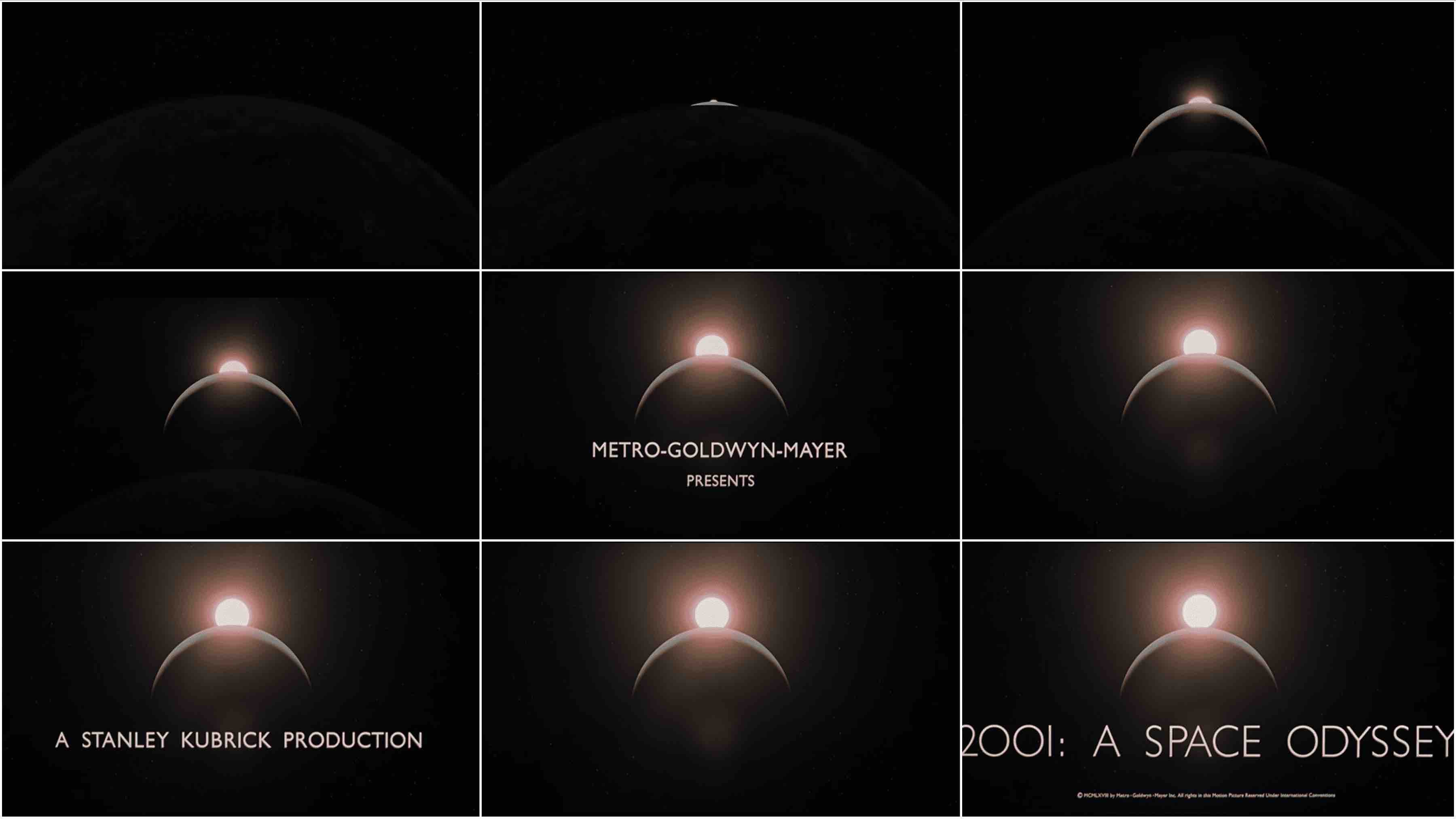}
  \caption{The opening sequence of \textit{2001: A Space Odyssey (1969)} using \textit{Creative Sound}~\cite{2001}.}
    \label{fig:space}
    \Description{This is the opening sequence of \textit{2001: A Space Odyssey (1969)} using \textit{Creative Sound}~\cite{2001}. The great opening of Kubrick's \textit{2001: A Space Odyssey (1969)}~\cite{2001} used a synchronized way. It played the opening portion of Richard Strauss's 1896 tone poem symphony to match the sun rising above the curvature of the earth and the moon as its high point of the sequence. Then there was also the non-synchronized way, in which images and sound operated somewhat independently.}
  \end{minipage}
  \hfill
  \begin{minipage}[b]{0.45\textwidth}
    \includegraphics[width=\linewidth]{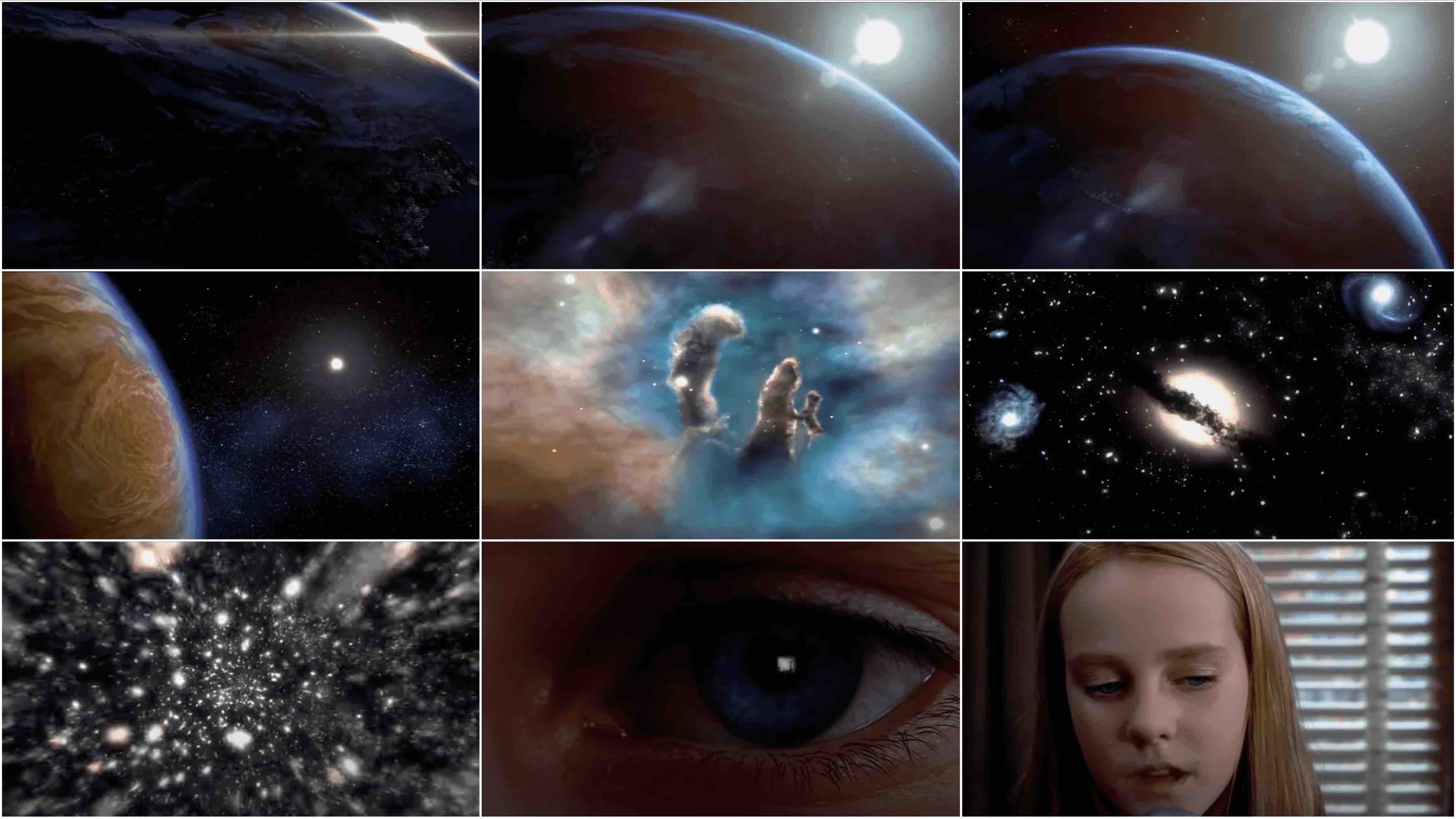}
  \caption{The opening sequence of \textit{Contact (1997)} using \textit{Creative Sound}~\cite{Contact}.}
  \label{fig:contact}
  \Description{This is the opening sequence of \textit{Contact (1997)} using \textit{Creative Sound}~\cite{Contact}. In the famous opening of Robert Zemeckis' \textit{Contact (1997)}~\cite{Contact} based on scientist Carl Sagan's vision of the universe, the sound traveled through the vacuum of space as radio transmission in radio waves.}
  \end{minipage}
\end{figure}

\begin{figure}[!ht]
  \centering
  \begin{minipage}[b]{0.45\textwidth}
  \includegraphics[width=\linewidth]{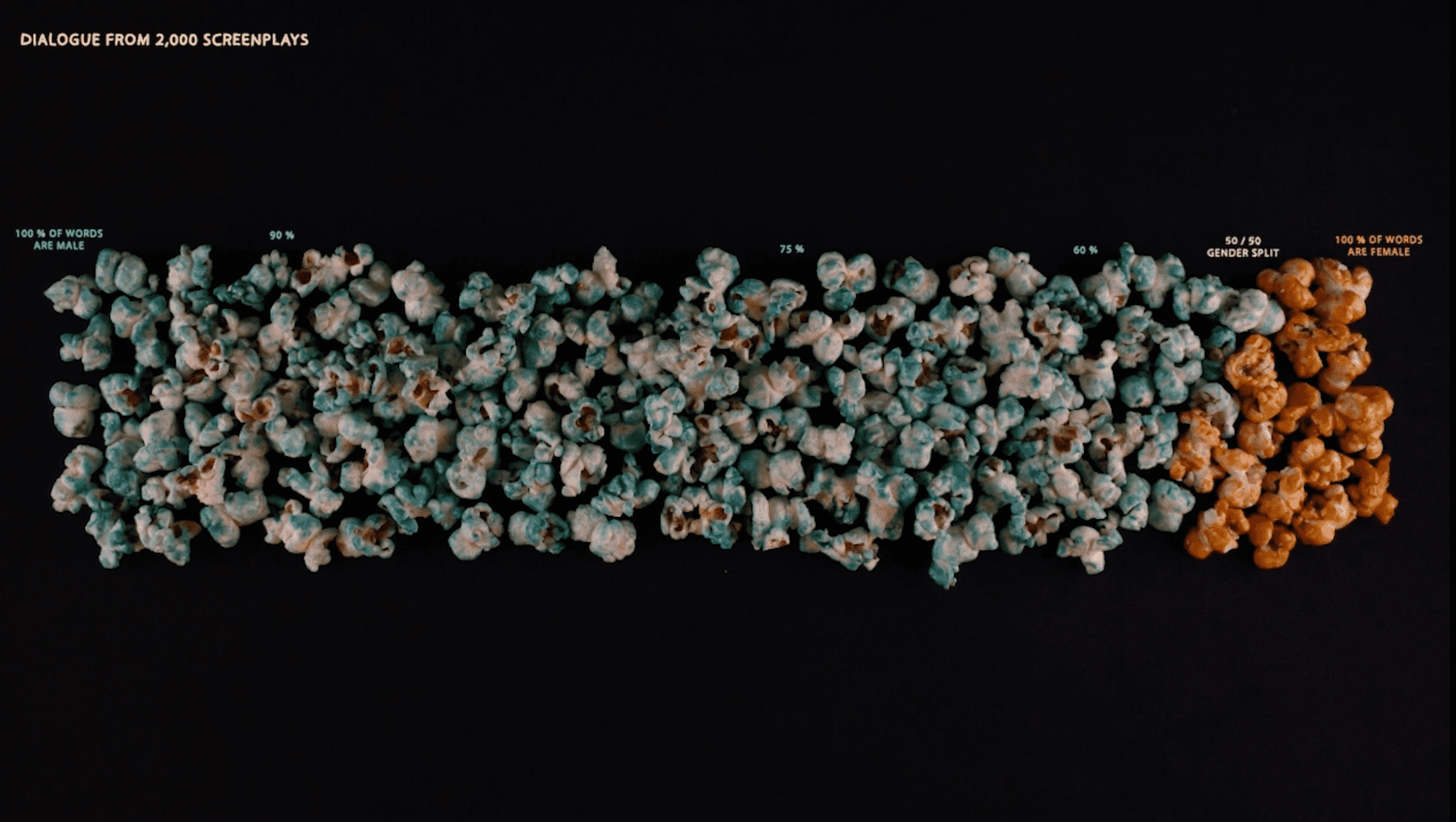}
  \caption{The image of the opening sequence of \textit{Film Dialogue by Gender} using \textit{Creative Sound}~\cite{FILM}.}
  \label{fig:dialogue}
  \Description{This is the opening sequence of \textit{Film Dialogue by Gender} using \textit{Creative Sound}~\cite{FILM}. The data video \textit{Film Dialogue by Gender (2019)}~\cite{FILM} used the sound of dialogue from films and the image of popcorn in the opening to echo the two important subjects of the story, namely, film and dialogue.}
  \end{minipage}
  \hfill
  \begin{minipage}[b]{0.45\textwidth}
    \includegraphics[width=\linewidth]{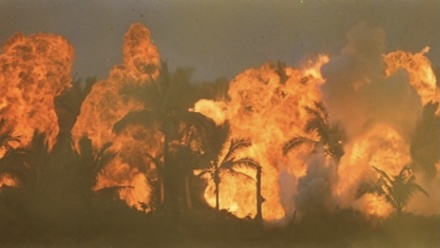}
  \caption{The explosion scene in the opening sequence of \textit{Apocalypse Now (1978)} using \textit{Big Bang}~\cite{Apocalypse}.}
  \Description{This is explosion scene in the opening sequence of \textit{Apocalypse Now (1978)} using \textit{Big Bang}~\cite{Apocalypse}. The famous opening sequence of Apocalypse Now (1978) used opening styles of symbolism, big bang and creative sound. It was big bang because it literally opened with a sudden explosion of the jungle to symbolize the bombing of Vietnam. It was also a good use of creative sound using low pitch sound effects of helicopter flying to demonize the effects of war destruction.}
  \end{minipage}
\end{figure}

\textbf{\creativesound}: This style uses sound and music as key elements to express the content, emotion, and rhythm of the opening sequence. Sound and image can not only synchronize together to create a consistent experience for the audience but also work against each other to build suspense and tension.
Many great examples of this opening style were accompanied by classical symphony and other great music for synchronized effects. The great opening of Kubrick's \textit{2001: A Space Odyssey (1969)}~\cite{2001} used synchronization. It played the opening portion of Richard Strauss's 1896 tone poem symphony to match the sun rising above the curvature of the earth and the moon as its high point of the sequence (see~\autoref{fig:space}). 
By contrast, in non-synchronization, images and sound operate independently. 
For example, in the opening of Robert Zemeckis' \textit{Contact (1997)}~\cite{Contact}, which is based on scientist Carl Sagan's vision of the universe, sound traveled through the vacuum of space as radio transmission in radio waves (see~\autoref{fig:contact}). As the camera traveled continuously farther from the earth, the older the radio transmissions the audience could hear until there was complete silence when the camera went too far beyond the observable universe.
These creative cinematic treatments of film opening sequences about the universe served to amplify the mysterious force of nature in a spectacular and respectful way.
The opening sequence of \textit{Apocalypse Now (1978)}~\cite{Apocalypse} was also a good use of creative sound that used low-pitched sound effects of a helicopter in flight to demonize the effects of war destruction.
As both film and data video are time-based, audio-visual art, the creative treatment of sound elements in films could be applied to data videos. In data videos, sound elements must be related to the topic or theme of the story, and they must complement the visual to convey the story topic. 
For example, the data video \textit{Film Dialogue by Gender (2019)}~\cite{FILM} used the dialogue from films and the image of popcorn (see~\autoref{fig:dialogue}) in the opening to echo the two important subjects of the story: film and dialogue.

\begin{figure}[!ht]
  \centering
  \begin{minipage}[b]{0.45\textwidth}
    \includegraphics[width=\linewidth]{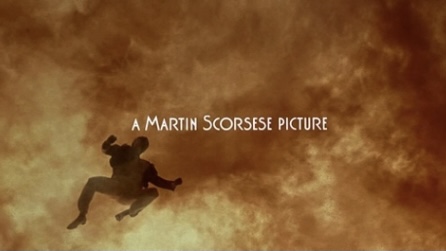}
  \caption{The explosion scene in the opening sequence of \textit{Casino} using \textit{Big Bang}~\cite{Casino}.}
  \label{fig:casino}
  \label{fig:apocalypse}
  \Description{This is the explosion scene in the opening sequence of \textit{Casino} using \textit{Big Bang}~\cite{Casino}. The opening of Martin Scorsese's \textit{Casino (1995)}~\cite{Casino} designed by Saul Bass started with a big bang of an explosion which was immediately followed by a surreal montage sequence made up of patterns of casino neon lights.}
  \end{minipage}
  \hfill
  \begin{minipage}[b]{0.45\textwidth}
    \includegraphics[width=\linewidth]{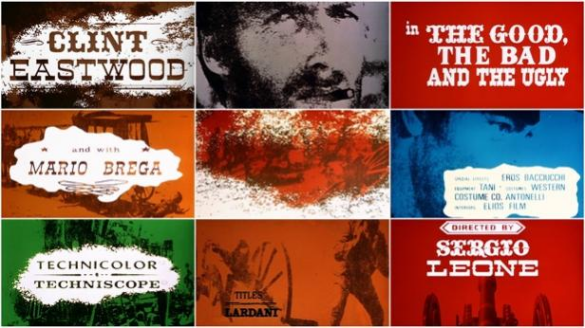}
  \caption{The opening sequence of \textit{The Good, the Bad and the Ugly (1966)} using \textit{Big Bang}~\cite{TheGood}.}
  \label{fig:the-good}
  \Description{This is the opening sequence of \textit{The Good, the Bad and the Ugly (1966)} using \textit{Big Bang}~\cite{TheGood}. It has a very energetic title sequence starting off with not just one but three bangs. Each bang represented each of the three film characters.}
  \end{minipage}
\end{figure}

\begin{figure}
  \centering
  \begin{minipage}[b]{0.45\textwidth}
    \includegraphics[width=\linewidth]{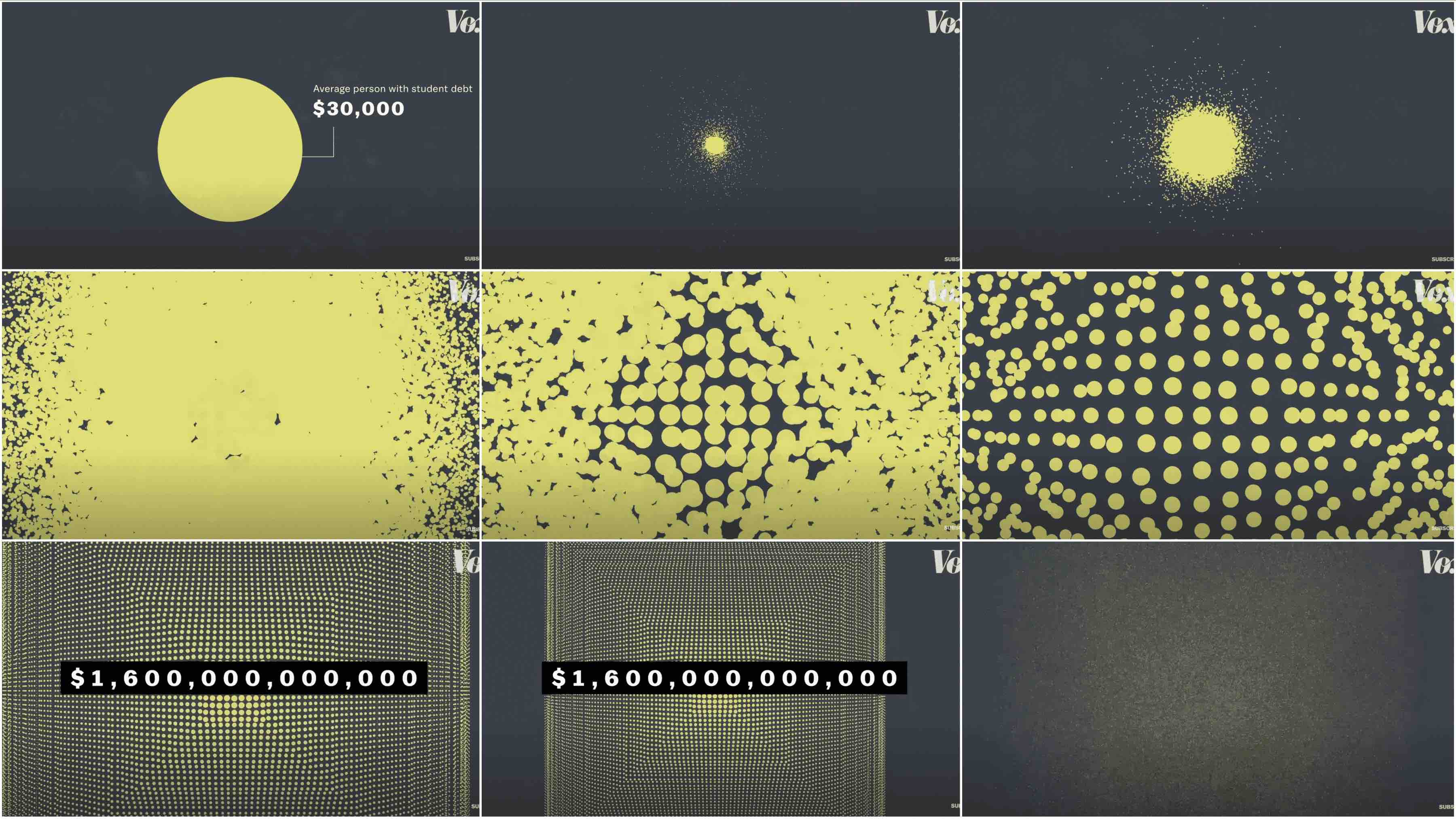}
  \caption{The opening sequence of \textit{All Student Debt in the US, Visualized} using \textit{Big Bang}~\cite{All}.}
  \label{fig:student-debt}
  \Description{This is the opening sequence of \textit{All Student Debt in the US, Visualized} using \textit{Big Bang}~\cite{All}. This data video \textit{All Student Debt in the US, Visualized (2019)}~\cite{All} was about the significant amount of all student debt Americans owe. In the opening, it used bubbles to represent each student's debt and a visual effect of one bubble expanding and exposing to generate thousands of bubbles to indicate the out-of-control situation.}
  \end{minipage}
  \hfill
  \begin{minipage}[b]{0.45\textwidth}
    \includegraphics[width=\linewidth]{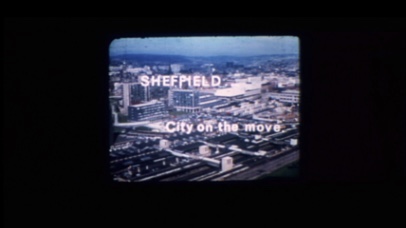}
  \caption{The opening of \textit{The Full Monty (1997)} using \textit{Old Footage}~\cite{TheFull}.}
  \label{fig:monty}
  \Description{This is the opening of \textit{The Full Monty (1997)} using \textit{Old Footage}~\cite{TheFull}. This film opened with old footage, the once-prosperous and energetic steel mills of Sheffield was now a town of the many unemployed where the main character, a former steelworker, was now stealing steel with his son in an old abandoned factory he once worked ironically.}
  \end{minipage}
\end{figure}

\textbf{\bigbang}: This style uses a sudden change of various elements such as color, composition, shape, size, or speed to create surprise, shock, and curiosity in the audience.
For example, the opening of \textit{Apocalypse Now (1978)}~\cite{Apocalypse}, which we referred to as using the \textit{Symbolism and Metaphor} style above, is also an example of \textit{Big Bang}. It literally opened with a sudden explosion of the jungle, as shown in ~\autoref{fig:apocalypse}.
A meaningful big bang in the opening should contextually relate to the following story content. 
For example, the opening of Martin Scorsese's \textit{Casino (1995)}~\cite{Casino} designed by Saul Bass started with a big bang of an explosion which was immediately followed by a surreal montage sequence made up of patterns of casino neon lights, as shown in~\autoref{fig:casino}. 
Another classic example of this category is the famous Western film \textit{The Good, The Bad and The Ugly (1966)}~\cite{TheGood}. Its opening shows a very energetic title sequence starting off with not just one but three bangs. Each bang represented each of the three film characters, as shown in ~\autoref{fig:the-good}.
In a data video, if the story content involves some chaos or out-of-control situations, the content has the potential to be visualized as an opening similar to a loud honk that signals for an important announcement to come.
One example is the data video \textit{All Student Debt in the US, Visualized (2019)}~\cite{All} that was about the significant amount of all student debt Americans owe. In the opening, bubbles were used to represent each student's debt. In the visual effect, one bubble expands and exposes thousands of bubbles to indicate the out-of-control situation, as shown in~\autoref{fig:student-debt}.

\begin{figure}[!ht]
  \centering
  \begin{minipage}[b]{0.45\textwidth}
    \includegraphics[width=\linewidth]{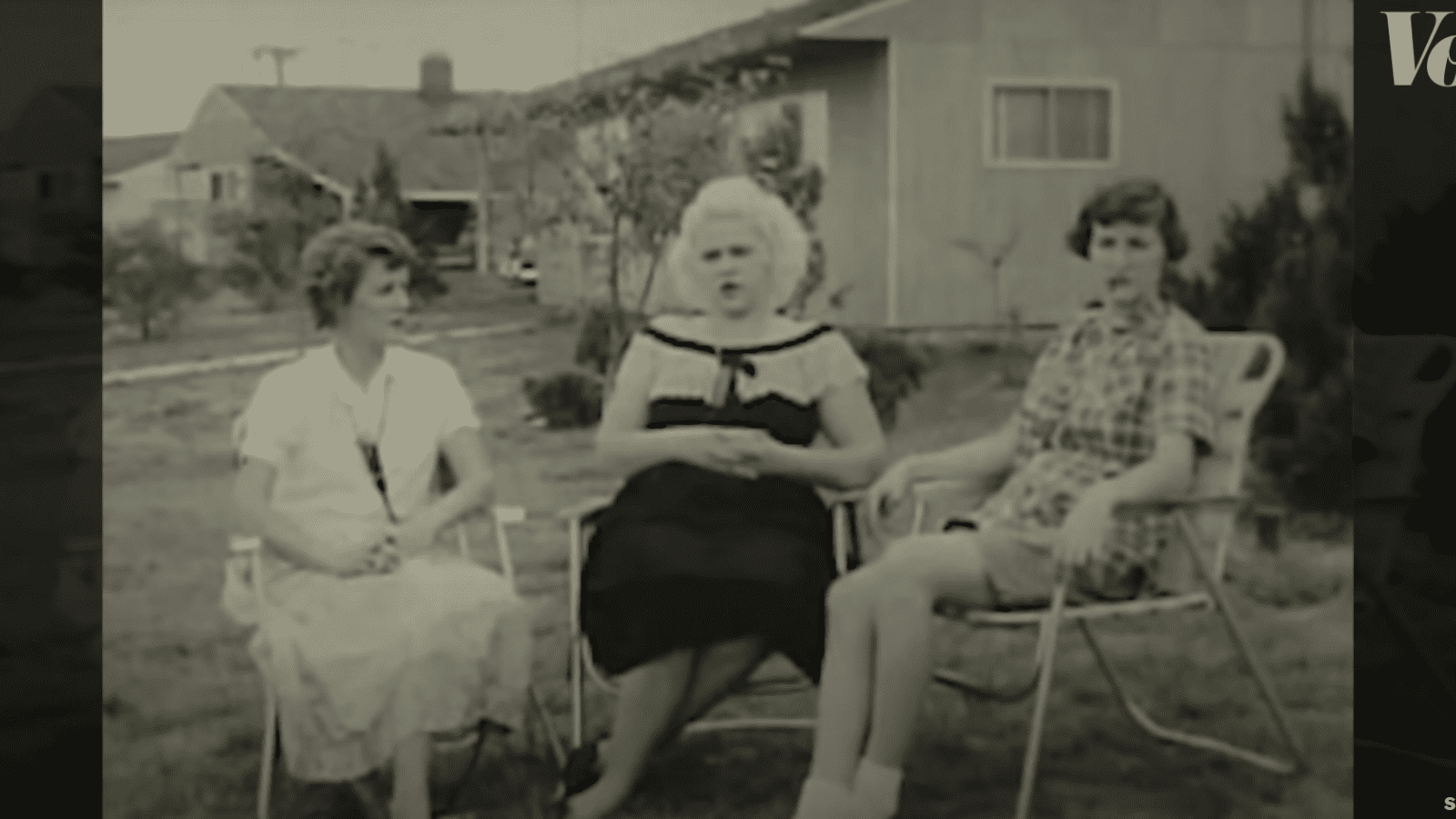}
  \caption{The opening of \textit{American Segregation, Mapped Day and Night (2019)} using \textit{Old Footage}~\cite{American}.}
  \label{fig:american}
   \Description{This is the opening of \textit{American Segregation, Mapped Day and Night (2019)} using \textit{Old Footage}~\cite{american}. It started with an old interview clip in which three housewives was talking about how they would feel unsafe to live in communities that is not all white, which was the opposite of racial harmony that the current society promotes.}
  \end{minipage}
  \hfill
  \begin{minipage}[b]{0.45\textwidth}
    \includegraphics[width=\linewidth]{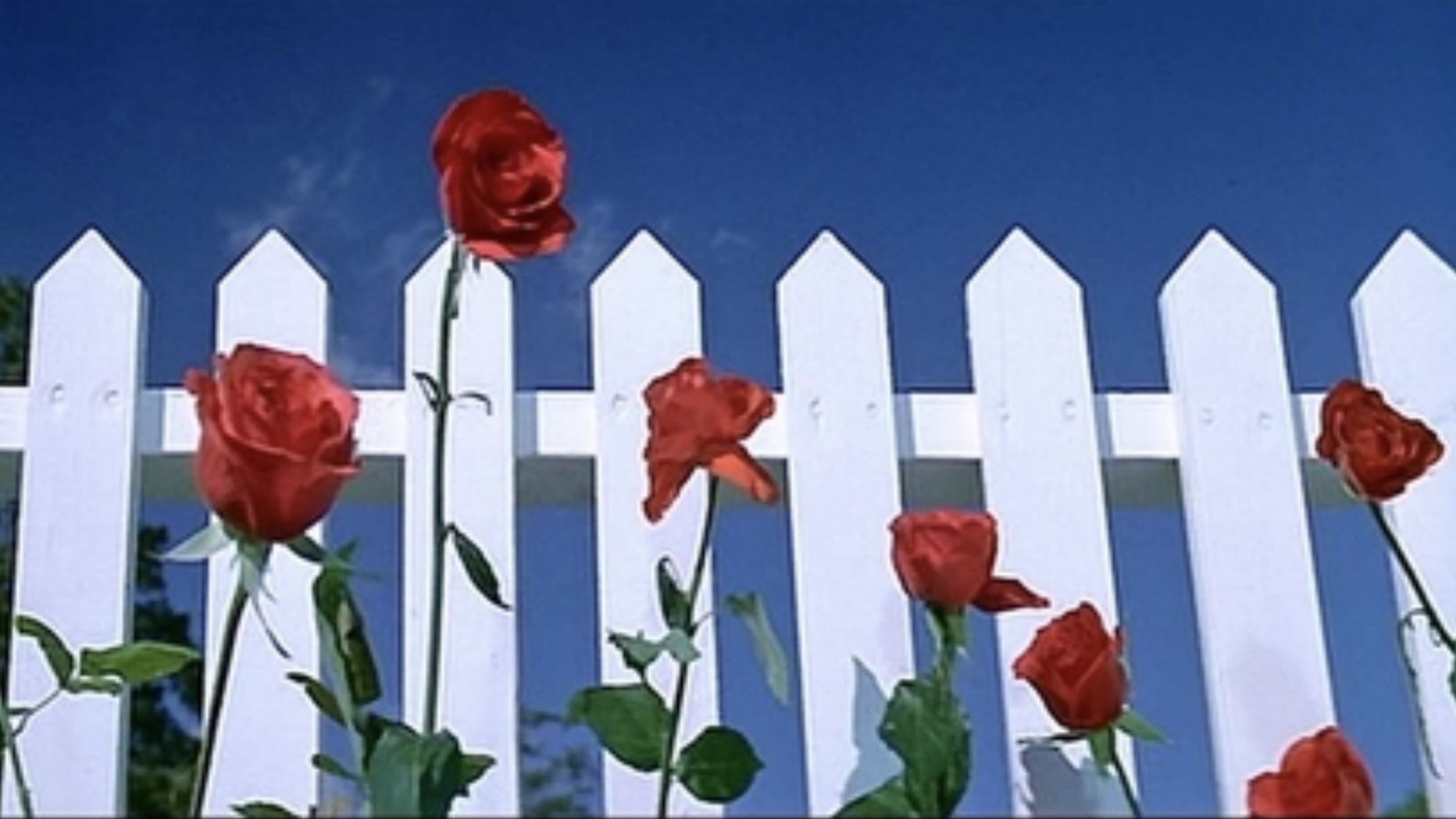}
  \caption{The opening of \textit{Blue Velvet (1986)} using \textit{Ending First}~\cite{Blue}.}
  \label{fig:blue}
   \Description{This is the opening of \textit{Blue Velvet (1986)} using \textit{Ending First}~\cite{Blue}. Each of the sequences echoes each other visually and thematically as the film took the audience back and forth from the different worlds of reality in the story.}
  \end{minipage}
\end{figure}

\textbf{\oldfootage}: This opening style uses historical image, photo, and/or soundbite to introduce the background of the story to draw the audience's interest to the then and now of the story, which should be almost always in opposition or in contrast to each other.
For example, the unique opening scene of Neill Blomkamp's \textit{District 9 (2009)}~\cite{District} used not only news footage but also BBC-type interview footage to open a Sci-Fi alien film on the topic of racial apartheid. 
When a story started with some old footage of the city or neighborhood of the story, it helped provide local context to the story. Examples of this type of opening can be found in Peter Cattaneo's \textit{The Full Monty (1997)}~\cite{TheFull}, Peter Jackson's \textit{Heavenly Creatures (1994)}~\cite{Heavenly}, and Martin Scorsese's \textit{Mean Streets (1973)}~\cite{mean}. These films opened with old footage of the glorious days of the town or the neighborhood where the story took place. They then showed the sharp contrast between the place's past and present as part of the story. The contrast in any form and content attracted attention. In \textit{The Full Monty (1997)}, as shown in~\autoref{fig:monty}, the once-prosperous and energetic steel mills of Sheffield was now a town of the many unemployed. The main character, a former steelworker, resorted to stealing steel with his son in an old abandoned factory where he once worked \st{ironically}.
In the opening sequence of \textit{Heavenly Creatures (1954)}~\cite{Heavenly}, right after New Zealand's Christchurch was introduced, the scene dramatically cut to a sinister murder plot in process with blood all over the female character's face. Another function of using old footage in an opening sequence in films such as \textit{Seven (1995)}~\cite{seven} and animation films such as \textit{Up (2009)}~\cite{up} and \textit{Minions (2015)}~\cite{Minions} is to show the backstory of the story. 
Similarly, many videos, especially those from news agencies, used old footage to show the background of the story. For example, the data video \textit{American Segregation, Mapped Day and Night}~\cite{American} started with an old interview clip in which three housewives were discussing how they would feel unsafe to live in communities that were not all white (see~\autoref{fig:american}), which was the opposite of racial harmony that the current society promotes.

\textbf{\myendingfirst}: This opening style shows part of the ending first to trigger the audience's curiosity about how the story leads to such an ending.
For example, in the famous poetic opening and ending sequences of David Lynch's \textit{Blue Velvet (1986)}~\cite{Blue}, as shown in~\autoref{fig:blue}, each of the sequences echoed each other visually and thematically as the film took the audience back and forth from the different worlds of reality in the story. As a motif of the film, the floating white feather that appeared in both the opening and ending sequences of \textit{Forrest Gump (1994)}~\cite{Forrest} thematically expressed the idea that life is full of ups and downs. The mid-air fueling of the warplanes in the opening sequence of \textit{Dr. Strangelove (1964)}~\cite{Strangelove} revealed that this opening was the beginning of the end when the ending of the film showed how the world ended in a nuclear holocaust.
This style of opening combines with the ending to form a complete loop of the story. When the data story tells an idea contrary to common belief or conventional wisdom, using \textit{Ending First} is suitable. For example, to grab the audience's attention, the data video \textit{Weed is not More Dangerous than Alcohol}~\cite{Weed} opened with a surprising finding: although marijuana was classified by the US federal government as one of the Schedule I drugs (the most dangerous drugs), the number of people who had died from marijuana overdose was actually zero. After that, the video went deeper into the topic---the advantages and disadvantages of legalizing marijuana. In the end, it looked back on the government policy about marijuana that was introduced in the opening.

%% file: sections/4-guideline.tex
\section{Guidelines}
Next, we introduce how we developed the guidelines through expert interview. Then, we give an overview of our guidelines.

\subsection{Expert Interview}
\subsubsection{Participants}
We invited eight experts, including two film researchers (E1 and E2 had over 15 and 10 years of experience in film practice and theory research), two directors (E3 had four years of experience in documentary film, and E4 had four years of experience in feature film), two cinematographers (E5 and E6 had four years and six years of experience in film photography), one film and television producer (E7 had over 15 years of experience in writing and directing film and television), and one film visual designer (E8 had over one year experience in film visual design). The interviews were conducted face-to-face or through online meetings. 

\begin{figure*}
  \includegraphics[width=0.9\textwidth]{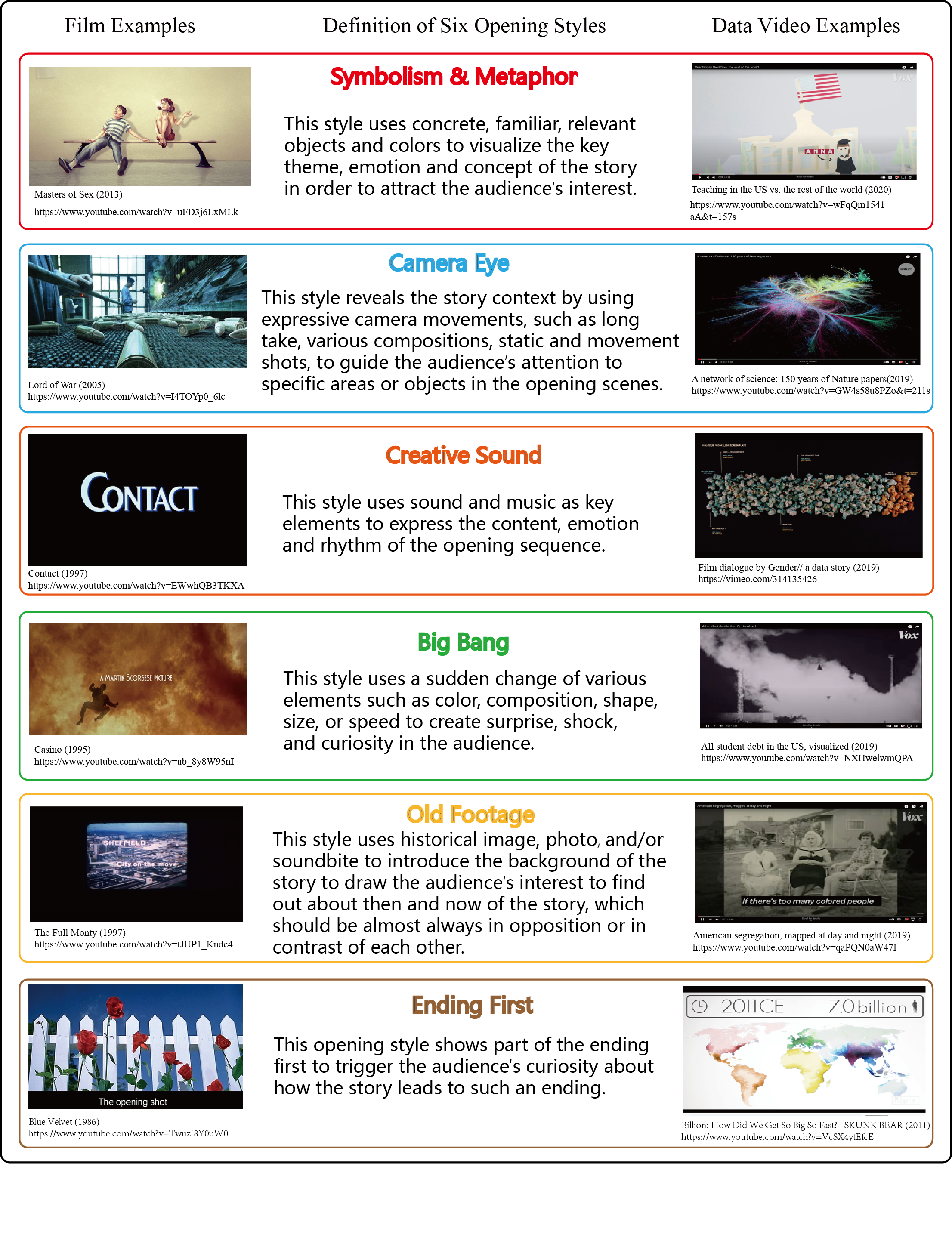}
  \caption{Films and data videos that were shown to participants in the expert interview.}
  \label{fig:interview-case}
  \Description{Illustration: Matching the Six Styles of Film Opening with Examples of Data Video Opening. Here are the films and data videos that we showed to participants in the expert interview.}
\end{figure*}

\subsubsection{Interview Procedure}
We first explained the purpose of the study as to help data video makers design attractive data video openings with cinematic styles. For experts who were not familiar with data videos, we showed them sample data videos (see~\autoref{fig:interview-case}). After the introduction, we asked experts to advise how data videos could better realize each cinematic opening style. We repeated the following procedure until all the six styles were discussed.
First, we introduced the definition and a film opening example of one cinematic opening style. Then we presented a data video that showed a related or adaptable approach in its opening. This was to inspire our experts on how the opening style could be realized in data videos. Moreover, their critical views on the data video could drive more insights on how data videos could be improved. We further asked three questions: (1) ``How well do you think the data video applied the cinematic opening style?'' (2) ``How can the data video be improved to better realize the cinematic opening style?'' and (3) ``Do you have any other pitfall or recommendation from your experience in film production that is also applicable in the context of data video?'' Overall, each interview lasted approximately for two hours.

\subsubsection{Interview Results Analysis}
The first and second authors analyzed the interview results. Using thematic analysis~\cite{braun2006using}, we first independently segmented, grouped, and labeled the feedback from experts to generate a set of mutually exclusive and instructive suggestions. We then had discussions to reach consensus. After that, the first, second, and third authors had several discussions and brainstorming by using the card sorting and affinity diagram methods to further refine and expand these suggestions to develop guidelines. Next, we demonstrate how the feedback from our experts affected our rationale behind the development of our guidelines for each cinematic opening style. 

\begin{table*}
\centering
  \begin{tabular}{|p{0.05\textwidth} | p{0.90\textwidth}|}
\hline
                      & \multicolumn{1}{c|}{{\textbf{\symbolandmetaphor}}}                                                                                            \\ 
\hline
G1.1                   & Select keywords from the voice-over script and look for related symbols and metaphors.                                            
\\ 
\hline
G1.2                   & Use simple symbols and metaphors.                                                        
\\ 
\hline
G1.3                   & Apply symbolism and metaphor not only to the story theme but also to how data can be represented.            \\ 
\hline
G1.4                   & Consider representing data through characters or human figures when appropriate.
\\ 
\hline
G1.5                   & Adapt classic fable or metaphor to data story to make an idea more universal.
\\ 
\hline
G1.6                   & Use parallel editing to show the actual and symbolic meaning of the data back and forth. \\ 
\hline
G1.7                   & Use colors not only for creating sensory responses but also for carrying information about time and space.
\\ 
\hline
                      & \multicolumn{1}{c|}{\textbf{\cameraeye}}                                                                                                       \\ 
\hline
G2.1                   & Achieve expressive composition through camera lens in close-up, medium or wide shot.   
\\ 
\hline
G2.2                   & Use symmetry to create harmony and asymmetry to create conflict and tension.           \\ 
\hline
G2.3                   & Use a series of long shots to create calmness and a series of quick shots to create excitement.    \\ 
\hline
G2.4                   & Hold the static shot long enough for the audience to digest the information. Keep the movement steady and consistent in speed.                  
\\ 
\hline
G2.5                   & Maintain the space and time continuity when using sequential shots.                                      \\ 
\hline
\multicolumn{1}{|c|}{} & \multicolumn{1}{c|}{\textbf{\creativesound}}                                                                                                   \\ 
\hline
G3.1                      & Use sound elements (voice-over recording, sound effects, and music) with different priorities for different communicative functions.                \\ 
\hline
G3.2                 & Use sound quality to reflect specific period of history.                                               
\\ 
\hline
G3.3                  & Handle sound elements (voice-over recording, sound effects, and music) separately in post-production.   
\\ 
\hline
G3.4                    & Let sound and image synchronize together to create consistent storytelling or let them work against each other for suspense and tension.
\\ 
\hline
G3.5                     & Use different treatments of sound elements such as volume and pitch to represent different characteristics of data.                 
\\ 
\hline
                      & \multicolumn{1}{c|}{\textbf{\bigbang}}                                                                                                          \\ 
\hline
G4.1                   & Relate the images of explosion in the opening to the theme and content of the story.                                                    
\\ 
\hline
G4.2                   & Consider moving story content that visualizes a chaotic situation to the opening.                      
\\ 
\hline
G4.3                   & Consider moving story content that involves sudden and rapid change of data to the opening.                    \\ 
\hline
G4.4                   & Use wide shot to show the full extent of change, expansion and development of the big bang in full view.  \\ 
\hline
G4.5                   & Consider visualizing extreme, huge contrasting or expanding data like an explosion as an opening style.                                            
\\ 
\hline
                      & \multicolumn{1}{c|}{\textbf{\oldfootage}}                                                   \\ 
\hline
G5.1                   & Keep the visual quality of old footage when such quality is unique and specific to its time.       
\\ 
\hline
G5.2                   & Select old footage that can show contrast between the past and present.                                    
\\ 
\hline
G5.3                   & Make sure the style of data visualization that is inserted into the old footage is aligned with the period of the old footage.                      
\\ 
\hline
                      & \multicolumn{1}{c|}{\textbf{\myendingfirst}}                                                                                                       \\ 
\hline
G6.1                   & Use Ending First for opening with unique visualization (the ``Wow'') followed by an explanation (the ``Why'').          
\\ 
\hline
G6.2                   & Use this style to show content that is contrary to common belief or conventional wisdom.                                      
\\ 
\hline
G6.3                   & Use similar elements from the ending in the opening but avoid making the opening identical to the ending.               
\\
\hline
\end{tabular}
\caption{\label{tab:guideline}Twenty-eight guidelines for applying the six cinematic opening styles to data videos.}
\end{table*}

\subsection{Overview of Guidelines}
Table~\ref{tab:guideline} presents all of our guidelines. Next, we combined our expert interview results to demonstrate the rationale behind the guidelines for each cinematic opening style.

\textbf{\symbolandmetaphor}: 
Almost all experts (6 out of 8) commented that the core challenge of applying this style is to make sure the audience can understand and relate to the symbols and metaphors and connect them with the story topic.
First, they gave practical suggestions on how to find understandable and related symbols and metaphors. G1.1, G1.3, G1.4, G1.5 and G1.7 provide such guidance according to the good practices that the experts and the authors saw from the films and data videos. For example, both E1 and E2 recommended that using familiar and interesting concepts (e.g., fables and classical stories) as symbols and metaphors, which corresponds to G1.5. 
Moreover, some experts gave advice on facilitating the understanding of the symbols and metaphors by manipulating their visual presentations. Specifically, E1 commented that \textit{``Symbols and metaphors as external content of the story should not confuse the audience; thus, `keep them simple' is extremely important like what we often say `simplicity is beautiful and less is more.' ''} which helped the development of G1.2. E7's feedback illustrated G1.6 well: \textit{``Using cross-cutting to show (back and forth) the corresponding parts between a symbol and what the symbol represents helps reveal their connections.''}

\textbf{\cameraeye}:
All experts agreed that camera movements are frequently used in film opening sequences to shape the viewers' perspectives. 
First, they emphasized how camera movements should be used to communicate emotions. Then the experts and the authors discussed applicable manipulations of camera movement to enhance emotional expression in the context of data videos and derived G2.1, G2.2, and G2.3. For example, G2.3 was developed on the basis of E1's suggestion that``\textit{shot duration determines the rhythm of the story, and a series of long shots often creates a calm feeling whereas a series of short shots creates excitement and intensity.''}
Moreover, as the data video presented a complex network visualization, it reminded experts that another important function of using camera movement was to enhance the audience's comprehension of the scenes. Correspondingly, we received several feedback on what principles the video should follow to introduce the visualization clearly (E5: \textit{``This three-dimensional tree is uncommon. The camera movement should maintain the consistency of story across both time and space, especially when the visualization is presented dynamically.''} E6 added, \textit{``The video used many jump cuts to present different parts of the visualization, which may be hard for the audience to understand. Normally, we use sequential shots to gradually change the distance between the camera and the scene, such as moving from long shot, to medium shot, and then to close-up shot, or the other way around''}).
After summarizing their feedback, the authors developed G2.4 and G2.5.

\textbf{\creativesound}:
The experts commented that a movie experience was only completed when both sound and visuals were presented, yet the importance of sound was often overlooked by many filmmakers.
The primary function of sound is to enhance the topic, plots, and emotions that are communicated by a story. Accordingly, the experts provided advice on how sound can help convey the story from different aspects. Overall, the experts recommended that sound could communicate the time of the story (E5: \textit{``The quality of sound can reflect some specific periods of history. For example, the noise of analog radio reminds people of old days''}), theme (E7: \textit{``The sound should match with the theme''}), and data patterns (E4: \textit{``The change of the sound volume could represent the increase or decrease of data, and different sound textures could represent data with different characteristics''}). These comments helped us develop G3.2, G3.4, and G3.5.  
Additionally, experts mentioned the complexity of sound in films and the need to avoid confusing audiences with too many sound elements. Experts suggested that sound elements, such as voice-over narration, sound effect, and music, serve different communicative and expressive functions with different priorities (E1, E2, E3), and that data video makers could handle them separately in post-production to prevent tangling them. We derived G3.1 and G3.3 from the above feedback.

\textbf{\bigbang}:
Big Bang is one way to move the cinematic climax to the beginning of a story to attract the audience's attention or create visual excitement, especially when movies involve explosion visual effects. Most of the experts (5 out of 8) suggested linking these explosive effects with data so that a data story can connect better with the audience. Specifically, E8 commented, ``\textit{Creating explosion effects is not difficult; the challenge is how to connect it with the story and the audience. Draw the audience's interest even before setting off the explosion so that when the explosion occurs, the audience can feel the strong impact.''} This feedback was reflected in G4.2, G4.4, and G4.5. E5 also suggested to designers that \textit{``The important point is why this data leads to collision or explosion. The image of explosion must be related to the content so that the result and the conflict can be revealed.''} This suggestion informed the development of G4.1 and G4.5.
Other experts suggested how \textit{Big Bang} should be visualized. G4.4 was illustrated in E1's comment that \textit{``In visualizing big bang effects, giving a full view is important so that the audience knows the spatial relation of the explosion. Don't show big bang effects in a series of close-up shots.''}
Moreover, some experts spoke about the narrative potential of \textit{Big Bang}. G4.3 was developed according to E7's suggestion that \textit{``On a metaphorical level, destructive and chaotic situation can be symbolized by this big bang effects.''}

\textbf{\oldfootage}:
The visual style of an image can hint different time and places to the audience. Experts suggested different ways in which relevant materials, both on content and visual, can be used in this style. 
First, on content, E2 agreed that old footage is an attractive opening: \textit{``Old footage can be used to show history's then-and-now connection in order to stimulate the audience's curiosity.''} This expert's view inspired us from the start to refine the definition of \textit{Old Footage} into the current version. G5.1 was illustrated in the documentary film director E3's comment: \textit{``Who uses this style must have a good understanding and sensibility of the historical context of the footage. They can consider how the history has impacted society and people of today as one approach or treatment of the footage.''}
Second, on visual, three participants mentioned examples of visual patterns. For example, E5 commented that \textit{``To show the relevance of the footage to a specific time, one can show its visual style according to its time. For example, the old footage of a train station in the 19th century was 16 frames per second, but the current projection speed is 24. As a result, there is some film speed effect that associates with the period of the footage.''} E8 agreed that \textit{``As time changes, the visual quality or style of the images changes also.''} After summarizing their comments, we derived G5.1, G5.2, and G5.3.

\textbf{\myendingfirst}:
Associating the opening and the ending is a common nonlinear narrative technique. G6.3 was illustrated in E2's comment: \textit{``Show the ending in the opening but make sure the real ending is not an entire repetition of what has been shown. It can be repeated with some variation with the goal to leave a strong impact to the audience.''} According to data stories, more specific suggestions were provided by experts. For example, E8 suggested that \textit{``This opening style is suitable to explain the unusual finding or something against conventional wisdom or common belief. The rest of the story can be about explaining why this is. The more the contrast between the opening and conventional belief, the better the storytelling is.''} E7 commented that \textit{``This nonlinear opening style lets the audience know the ending first without spoiling the ending and keeps the audience interested in knowing more. The focus of the story is how the ending happened.''} Accordingly, G6.2 was developed according to E8's suggestion, and G6.1 was illustrated in E7's comment.

%% file: sections/5-evaluation.tex
\section{Evaluation}
We assess how well our design guidelines can help users design attractive data video openings by considering two questions: (1) whether users can understand and apply the design guidelines \st{(usability)}, and (2) whether the design guidelines can \rv{help achieve cinematic style openings for data videos and improve their attractiveness}. 
\st{To answer these questions, we evaluated the usability and effectiveness of our guidelines through a comparative study. First, two groups of participants designed data video openings with and without our guidelines, respectively, which are then evaluated by experts from the film industry and the general public.} 
\rv{To answer the first question, we asked participants in our workshop to rate the overall clarity and usability of our teaching materials for each cinematic style, as well as the clarity and usefulness of each guideline in our workshop. 
As for the second question, we conducted a comparative study to ask the general public and experts from the film industry to rate data video opening designs by participants with and without our guidelines in the workshop according to different metrics. Specifically, the experts rated the \textit{creativeness} and \textit{cinema} of those designs. Additionally, both the general public and experts rated the \textit{attractiveness} of those designs. We selected these metrics for the following two reasons. 
First, to know whether our guidelines can help improve data video openings from the perspective of cinematography, we decided on the metrics creativeness and cinematic expression that are widely used to evaluate the cinematic styles of a film according to feedback from our collaborated film experts. Second, while the goal of a data video in its entirety may be to convey insights in data to its audience and thus memorability and understandability are reasonably metrics, the goal of the opening of a data video, we argue, is to capture its audiences' attention by attracting them to continue watching the rest of it. Hence, we decided on the metric attractiveness.}
Next, we illustrate our study design in detail.

\begin{figure*}[!ht]
  \includegraphics[width=\textwidth]{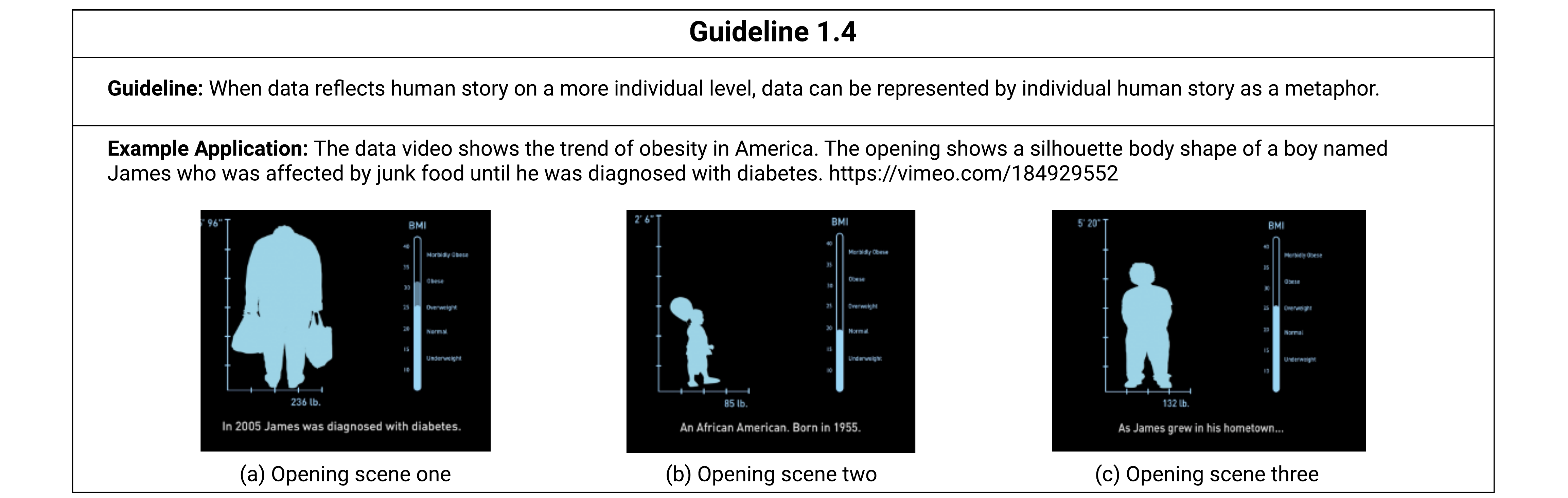}
  \caption{In the workshop, we provided participants with examples of each guideline. This figure shows the example of G1.4.}
  \label{fig:guidline-example}
  \Description{This is a figure shows the example of G1.4. In the workshop, we provided participants with examples of each guideline.}
\end{figure*}

\begin{figure*}
  \includegraphics[width=0.9\textwidth]{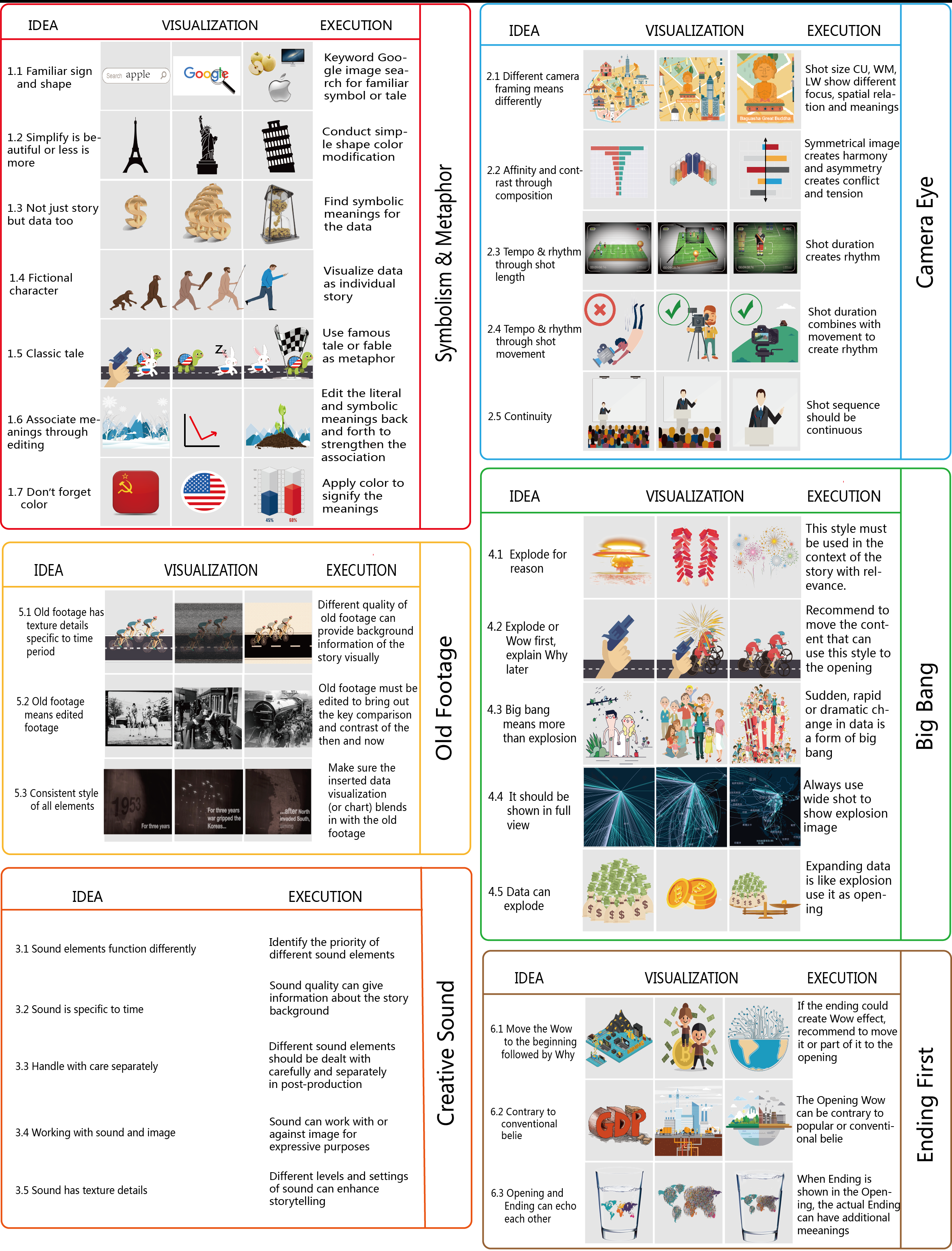}
  \caption{Overview of our guidelines including their core ideas, execution tips, and illustrations.}
  \label{fig:guidelines}
   \Description{Here is overview of our 28 guidelines including their core ideas, execution tips, and illustrations shown to the participants in our workshop.}
\end{figure*}

\subsection{Workshop}
We first conducted a workshop for collecting data video openings designed by participants with and without our design guidelines. The workshop was also for gathering qualitative feedback on the usability and inspiration of our guidelines. 

\subsubsection{Participants}
We recruited 20 participants by advertising on online social platforms and snowball sampling. Demographically, 11 were female and 9 were male, and their ages were between 24 and 28. Fourteen participants were data visualization researchers, and 6 were data analysts from various industries, including adverting, healthcare, finance, and e-commerce. All participants had experiences and needs of data storytelling. They applied data storytelling in presentations to report their data analysis results, to give a pitching proposal, and gave examples when they instructed domain experts on using the visual analytics systems they developed.

\subsubsection{Study and Teaching Materials}
During the workshop, we provided participants with two datasets related to two topics of general interest: \textit{The top 10 causes of death} and \textit{The global trends in overweight and obesity}. Following the method of previous work~\cite{bach2018design,amini2015understanding}, we gave some data facts (e.g., trend, rank, and extreme~\cite{wang2019datashot}) and corresponding data visualizations extracted from the datasets for participants to select from. This was to focus participants on storytelling instead of data exploration. 
We also provided participants with teaching materials to facilitate the learning process. Specifically, for each guideline, we provided its definition with an application example, as shown in~\autoref{fig:guidline-example}. We further explained it from three aspects, namely, its core idea, execution tips, and visual illustration, as shown in ~\autoref{fig:guidelines}.
Finally, we provided a template table for participants to document the outlines of their stories and the designs of the openings in the form of a storyboard, as shown in~\autoref{fig:storyboard-example}. All related materials are provided in the supplementary material.

\subsubsection{Study Design and Procedure}
We randomly and evenly assigned our participants into two groups. Each group had seven researchers in data visualization and three data analysts. We referred to the group who would use our guidelines as \emph{guideline participants}, and the other group who would not use our guidelines as \emph{non-guideline participants} in the remainder of this paper. We then randomly and evenly assigned participants from each group to use one of the datasets. Due to the COVID-19 pandemic, participants joined our study through online meetings, and they either used drawing tools on their laptops or tablets or sketched on papers and took photos to present their designs.

Our workshop had three phases: (1) \emph{introduction phase}, (2) \emph{design phase}, and (3) \emph{feedback phase}. In the \emph{introduction phase}, for \textit{guideline participants}, we introduced the workshop procedure, guidelines, datasets, and concepts of data video opening and storyboard. This phase took around 15 minutes. For \textit{non-guideline participants}, we introduced the same things except the guidelines, and this phase took around 10 minutes. 
In the \emph{design phase}, all participants were given 15 minutes to explore the dataset and decide their stories' titles and outlines. Then, they were given 40 minutes to finish their storyboards to show their designs of the openings of their stories. \textit{Guideline participants} had extra 20 minutes to learn the guidelines. Participants could write down their design ideas if they had difficulty demonstrating their visual designs through sketching. Finally, every participant introduced their designs.
In the \emph{feedback phase}, we invited \textit{guideline participants} to rate the attractiveness, clarity, and usability (to what extent they could apply the style by using our teaching materials) of each cinematic opening style, as well as the clarity and usefulness (to what extent the guideline could help them achieve the style) of each guideline using a 7-point Likert scale. Afterward, we interviewed them to understand their design considerations and their comments on the opening styles and guidelines. The interview questions were centered around: (1) which opening styles they did not know or had little knowledge about how to achieve them before our workshop; (2) which opening styles they regarded as (in)effective in attracting audiences and why; (3) which opening styles and guidelines they used in our workshop and out of what considerations; (4) why they excluded alternative opening styles; (5) which opening styles and guidelines they found unclear and why; (6) what challenges they encountered when designing the opening; and (7) what suggestions they had on our guidelines. 
As for the \textit{non-guideline participants}, we first interviewed them to understand their design considerations and the challenges they encountered during the design. We then asked them to read our guidelines and asked whether, and if so, how their designs could be improved by using our guidelines.
Each interview took around 10-15 minutes and was audio-recorded for subsequent analysis.  

\begin{figure*}
  \includegraphics[width=0.9\textwidth]{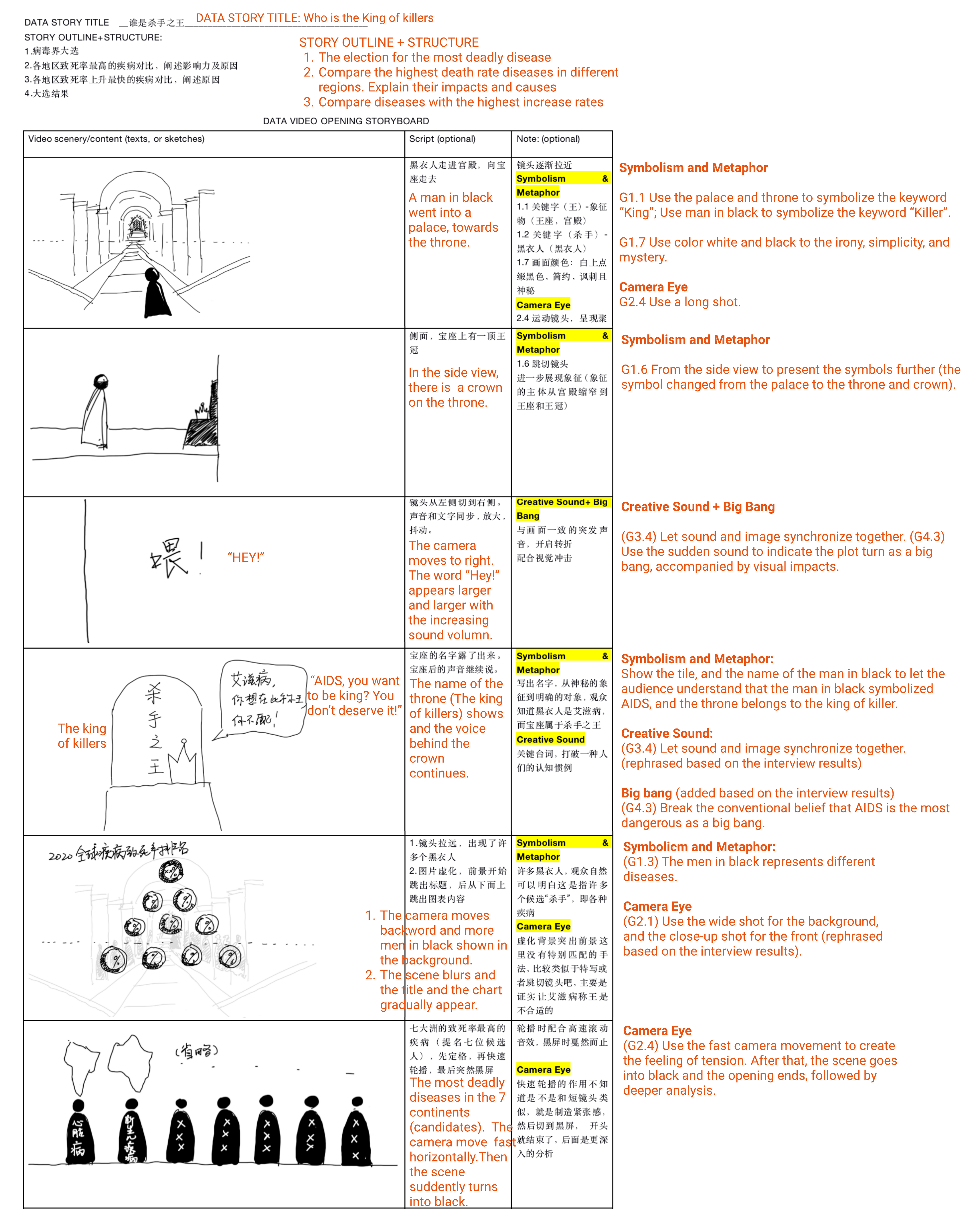}
  \caption{This figure is an example of the storyboard created by a \textit{guideline participant}. The top part is the story's title and outline. The lower table is the storyboard for the opening. Each row represents one key frame. The columns from left to right show the visual design, the description of the design, and the note indicating the used styles and guidelines. The orange text is the translation from the authors, which is based on both the storyboard and the interview with the participant.}
  \label{fig:storyboard-example}
  \Description{This figure is an example of the storyboard created by a guideline participant. The top part is the story's title and outline. The lower table is the storyboard for the opening. Each row represents one key frame. The columns from left to right show the visual design, the description of the design, and the note indicating the used styles and guidelines. The orange text is the translation from the authors, which is based on both the storyboard and the interview with the participant}
\end{figure*}

\subsection{Comparative Study}
To understand whether those story openings designed with our guidelines were truly more \st{effective in attracting} \rv{attractive to} audiences than those designed without our guidelines, we further recruited experts from the film industry and the general public to evaluate the storyboards from our workshop.

\subsubsection{Participants}
We recruited three experts from the film industry and different universities: a film practitioner with over 20 years of experience in industry and academia, a film screenwriter and director with over 10 years of experience in industry and academia, and an animation artist with over 8 years of experience in industry and academia.
We also recruited 20 general participants through advertising on social media platforms and snowball sampling. 

\subsubsection{Study Design and Procedure}
Both the experts and general participants rated all 20 storyboards, which were presented randomly. They were told that these storyboards were homework of students in a data storytelling class and did not know about guidelines. 
The experts rated the creativeness, attractiveness, and cinematic expression (i.e., to what extent these storyboards could be regarded as being cinematic) of the storyboards using a 7-point Likert scale. 
The general participants rated the attractiveness of each storyboard by considering how much they would like to continue to watch the story according to the storyboard.
To make sure that the experts and general participants could understand the storyboards, we provided them with the title and outline of each story. If a participant could not understand a storyboard in 2 minutes, then they could skip rating it. Furthermore, we emphasized that they should rate on the basis of design ideas of the storyboards instead of the effectiveness of the story titles and outlines.

%% file: sections/6-results.tex
\section{Results}
In this section, we report our evaluation results. In the first subsection, we report the feedback from the \textit{guideline participants} and \textit{non-guideline participants}. In the next subsection, we report quantitative results about the evaluation of the storyboards by experts and the general public.

\begin{figure*}[!ht]
  \includegraphics[width=\textwidth]{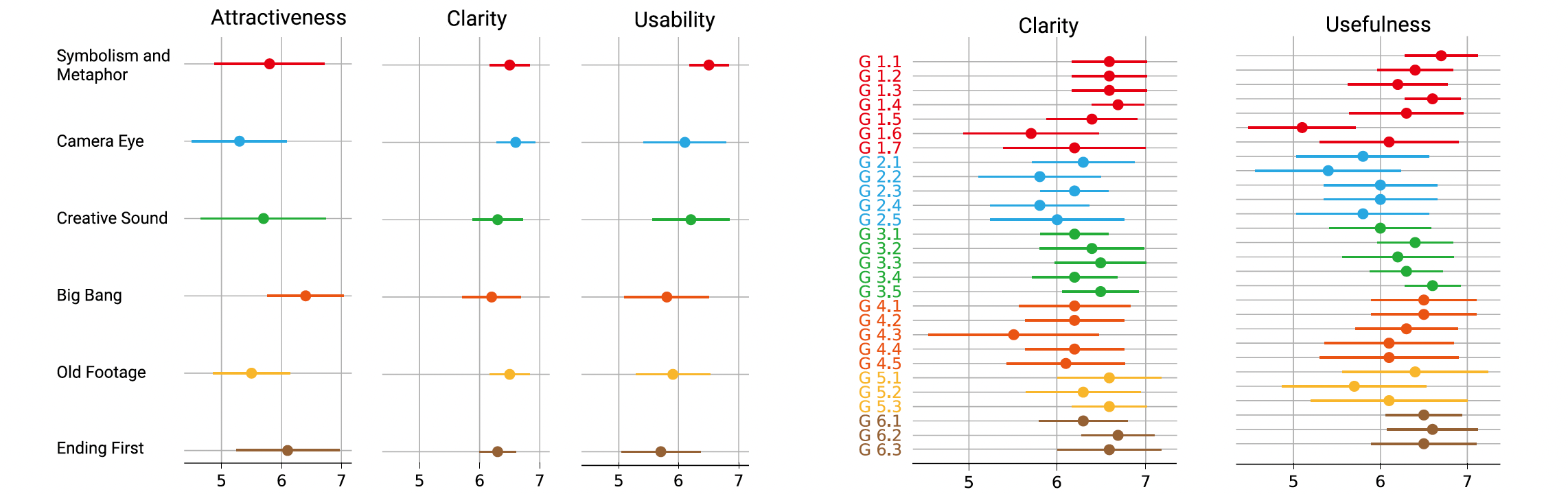}
  \caption{\textit{Guideline participants}' rating with 95\% confidence intervals on the cinematic opening styles and guidelines.}
  \Description{This figure shows the \textit{guideline participants}' ratings on the cinematic opening styles and guidelines. For each style, the specific average scores include: the attractiveness of Symbolism and Metaphor ($Mean = 5.80, SD = 1.47$), Camera Eye ($Mean = 5.30, SD = 1.27$), Creative Sound ($Mean = 5.70, SD = 1.68$), Big Bang ($Mean = 6.40, SD = 1.02$), Old Footage ($Mean = 5.50, SD = 1.02$), and Ending First ($Mean = 6.10, SD = 1.37$), the clarity of Symbolism and Metaphor ($Mean = 6.50, SD = 0.53$), Camera Eye ($Mean = 6.60, SD = 0.52$), Creative Sound ($Mean = 6.30, SD = 0.67$), Big Bang ($Mean = 6.20, SD = 0.79$), Old Footage ($Mean = 6.50, SD = 0.53$), and Ending First ($Mean = 6.30, SD = 0.48$), and the usability of Symbolism and Metaphor ($Mean = 6.50, SD = 0.53$), Camera Eye ($Mean = 6.10, SD = 1.10$), Creative Sound ($Mean = 6.20, SD = 1.03$), Big Bang ($Mean = 5.80, SD = 1.14$), Old Footage ($Mean = 5.90, SD = 0.99$), and Ending First ($Mean = 5.70, SD = 1.06$). 
  For each guideline, the specific average scores include: (1) the clarity of G1.1 ($Mean = 6.60, SD = 0.70$), G1.2 ($Mean = 6.60, SD = 0.70$), G1.3 ($Mean = 6.60, SD = 0.70$), G1.4 ($Mean = 6.70, SD = 0.48$), G1.5 ($Mean = 6.40, SD = 0.84$), G1.6 ($Mean = 5.70, SD = 1.25$), G1.7 ($Mean = 6.20, SD = 1.32$), G2.1 ($Mean = 6.30, SD = 0.95$), G2.2 ($Mean = 5.80, SD =1.14 $), G2.3 ($Mean = 6.20, SD = 0.63$), G2.4 ($Mean = 5.80, SD = 0.92$), G2.5 ($Mean = 6.00, SD = 1.25$), G3.1 ($Mean = 6.20, SD = 0.63$), G3.2 ($Mean = 6.40, SD = 0.97$), G3.3 ($Mean = 6.50, SD = 0.85$), G3.4 ($Mean = 6.20, SD = 0.79$), G3.5 ($Mean = 6.50, SD = 0.71$), G4.1 ($Mean = 6.20, SD = 1.03$), G4.2 ($Mean = 6.20, SD = 0.92$), G4.3 ($Mean = 5.50, SD = 1.58$), G4.4 ($Mean = 6.20, SD = 0.92$), G4.5 ($Mean = 6.10, SD = 1.10$), G5.1 ($Mean = 6.60, SD = 0.97$), G5.2 ($Mean = 6.30, SD = 1.06$), G5.3 ($Mean = 6.60, SD = 0.70$), G6.1 ($Mean = 6.30, SD = 0.82$), G6.2 ($Mean = 6.70, SD = 0.67$), G6.3 ($Mean = 6.60, SD = 0.97$), and (2) the usefulness of G1.1 ($Mean = 6.70, SD = 0.67$), G1.2 ($Mean = 6.40, SD = 0.70$), G1.3 ($Mean = 6.20, SD = 0.92$), G1.4 ($Mean = 6.60, SD = 0.52$), G1.5 ($Mean = 6.30, SD = 1.06$), G1.6 ($Mean = 5.10, SD = 0.99$), G1.7 ($Mean = 6.10, SD = 1.29$), G2.1 ($Mean = 5.80, SD = 1.23$), G2.2 ($Mean = 5.40, SD = 1.35$), G2.3 ($Mean = 6.00, SD = 1.05$), G2.4 ($Mean = 6.00, SD = 1.05$), G2.5 ($Mean = 5.80, SD = 1.23$), G3.1 ($Mean = 6.00, SD = 0.94$), G3.2 ($Mean = 6.40, SD = 0.70$), G3.3 ($Mean = 6.20, SD = 1.03$), G3.4 ($Mean = 6.30, SD = 0.67$), G3.5 ($Mean = 6.60, SD = 0.52$), G4.1 ($Mean = 6.50, SD = 0.97$), G4.2 ($Mean = 6.50, SD = 0.97$), G4.3 ($Mean = 6.30, SD = 0.95$), G4.4 ($Mean = 6.10, SD = 1.20$), G4.5 ($Mean = 6.10, SD = 1.29$), G5.1 ($Mean = 6.40, SD = 1.35$), G5.2 ($Mean = 5.70, SD = 1.34$), G5.3 ($Mean = 6.10, SD = 1.45$), G6.1 ($Mean = 6.50, SD = 0.71$), G6.2 ($Mean = 6.60, SD = 0.84$), G6.3 ($Mean = 6.50, SD = 0.97$).}
  \label{fig:workshop-score}
\end{figure*}

\subsection{Workshop Results}
\rv{In this subsection, we analyze the interviews and questionnaire feedback from participants who did and did not use guidelines.}
\subsubsection{Guideline Participants' Feedback about Using the Guidelines to Design Data Video Openings}
\label{sec:workshop-result}
We analyzed the labels of \textit{guideline participants} (P1-P10) about the styles and guidelines that they used. All six styles were applied in the following frequencies: \textit{Symbolism and Metaphor (7)}, \textit{Camera Eye (5)}, \textit{Creative Sound (4)}, \textit{Old Footage (4)}, \textit{Big Bang (3)}, and \textit{Ending First (1)}. Most guidelines for each style were applied, and most \textit{guideline participants} (7 out of 10) combined more than one style in their opening. Specifically, 3 participants used 2 styles, 1 participant used 3 styles, 2 participants used 4 styles, and 1 participant used all the styles. 
All participants reported that the six cinematic opening styles were comprehensive, and they did not recommend any other style to be added.
Overall, participants indicated that \textit{our styles and guidelines inspired their designs} \inlinegraphics{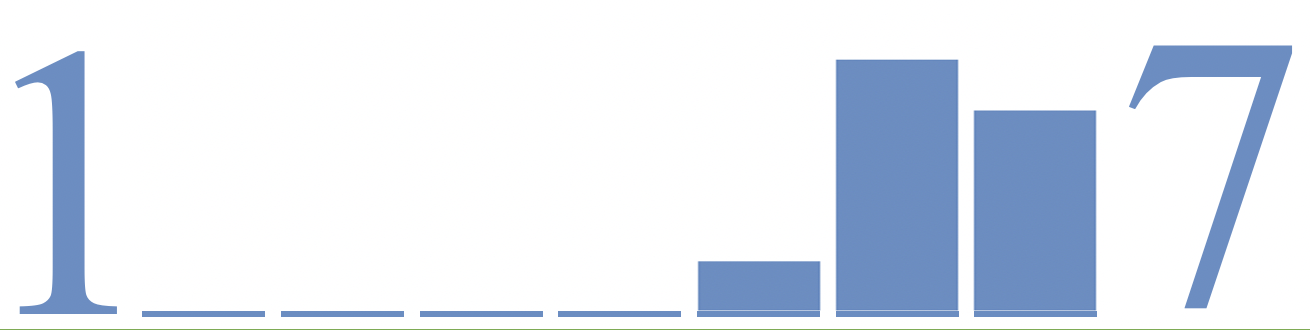} (M = 6.3, SD = 0.67), and \textit{they were satisfied with their final designs} \inlinegraphics{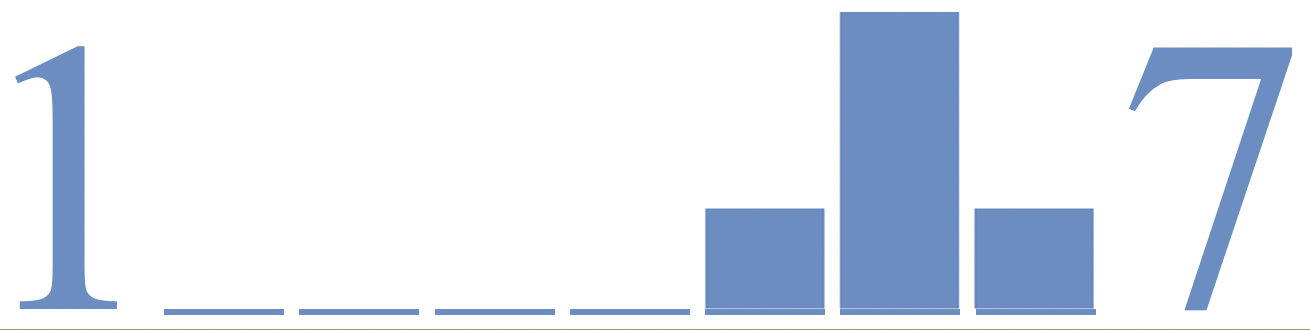} (M = 6, SD = 0.67) on a 7-point Likert scale.
With respect to the attractiveness, clarity, and usability (to what extent they could apply the style by using our teaching materials) of each cinematic opening style, as well as the clarity and usefulness (to what extent the guideline could help them achieve the style), the participants also gave positive scores (see~\autoref{fig:workshop-score}).

During the interview, we asked \textit{guideline participants} which styles and guidelines they found effective in attracting the audiences and why. Seven out of 10 \textit{guideline participants} thought all these styles were effective in attracting the audience (\textit{P4: ``They are good for different stories and purposes. Ending First is suitable for data videos that intend to explain something to make the audience curious. Big Bang has an impactful visual effect that surprises the audience;''} \textit{P3: ``I would like to try all of them''}).
They especially appreciated styles and guidelines that were new to them or were rare in data videos. For example, P1 commented that the style \textit{Creative Sound} was often seen in film previews but rarely in data videos, and it can enhance the creativeness of data videos. Seven out of 10 participants emphasized that this style was helpful in affecting the moods and emotions of the audience. P9 noted \textit{``The guidelines provided many sophisticated treatments of sound elements. A video applying them can demonstrate the efforts of the video maker and impress the audience.''} Similarly, \textit{Ending First} and \textit{Camera Eye} were praised for their creativeness (P1:\textit{``I didn't like using 'Ending First' before, but guideline 6.2 provided me with a new way to achieve it by showing something contrary to common belief, and I feel it could surprise the audiences;'' P2: \textit{``The long shot is good. It is used more and more often by televisions but it is relatively new to data videos''}}).



\subsubsection{Non-guideline Participants' Feedback about Designing Data Video Openings}
We asked non-guideline participants (P11-20) what challenges they had encountered during the workshop. Subsequently, they were asked to read our teaching materials and report whether, and if so, how they would change their opening designs according to our styles and guidelines. The challenges they faced could be classified into two aspects, which could be alleviated by our styles and guidelines.

The first challenge was how to logically transition from the opening to the main story (P11: \textit{``There is a gap between the opening and the main story''}). Two participants found that the style \textit{Symbolism and Metaphor} was quite helpful by connecting the opening with the story topic symbolically or metaphorically (P11: \textit{``Guideline 1.4 helps by recommending starting from an individual story to introduce a large picture''}; P12: \textit{``Both Guidelines 1.4 and 1.5 provided new ideas for me to design the story''). Moreover, P16 noted \textit{``If I had seen the material early, I could have more information about how to design the story flow.''}}
The second challenge was how to make the visual effects compelling (P15: \textit{``It is hard to imagine what the final visual will look like;''} P13: \textit{``My story is about Africa, but I don't know how to indicate that in the opening. Should I use scenes from movies about Africa?''}). Many participants gave new design ideas after reading our styles and guidelines. For example, P13 said: \textit{``I can use a long shot to show small regions first and then the whole Africa as an opening (Guideline 2.2). I can also use old footage to show the scenes in Africa. Those styles and guidelines are really inspiring.''} P16 commented that the teaching materials \textit{``let me learn a lot. The guidelines for the style Camera Eye are really helpful as they also followed traditions in presenting data visualizations such as guidelines 2.1, 2.3, and 2.5''.}

\begin{figure*}
  \includegraphics[width=0.5\textwidth]{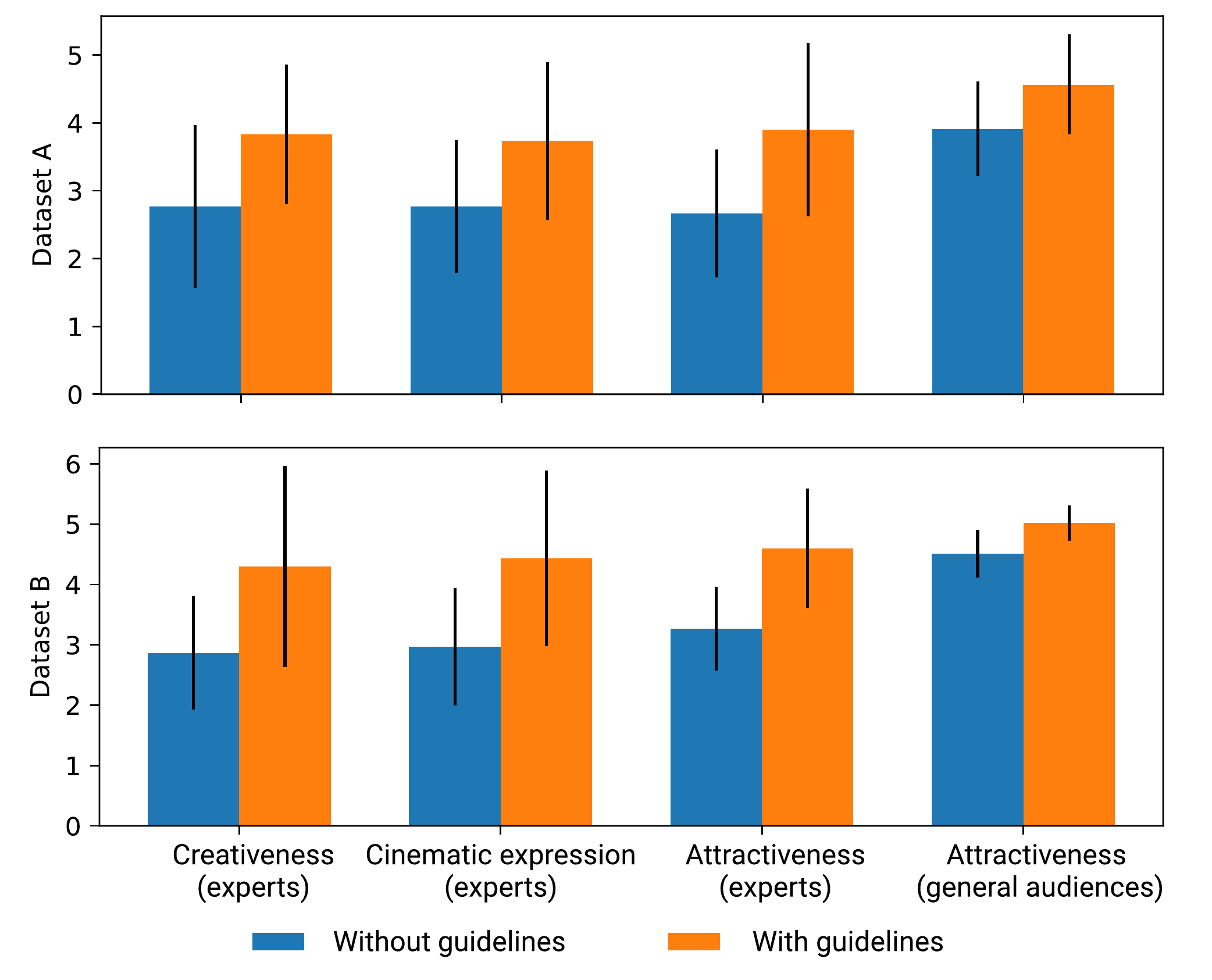}
  \Description[title]{detail description}
  \caption{Experts and the general audiences' ratings on the storyboards \rv{with 95\% confidence interval}. The upper chart shows the scores of storyboards using the dataset \textit{Top 10 causes of death}, and the lower chart shows the scores of storyboards using the dataset \textit{The global trends in overweight and obesity.}}
  \Description{This figure shows the experts and general audiences' rating on storyboards with and without our guidelines. For storyboards about the dataset \textit{Top ten causes of death}, the specific average scores include: (1) the creativeness of the storyboards using the guidelines ($Mean = 3.83, SD = 0.91$) and the storyboards not using the guidelines ($Mean = 2.77, SD = 1.06$), (2) the cinema of the storyboards using the guidelines ($Mean = 3.73, SD = 1.03$) and the storyboards not using the guidelines ($Mean = 2.77, SD = 0.86$), (3) the attractiveness (from experts) of the storyboards using the guidelines ($Mean = 3.90 SD = 1.13 $) and the storyboards not using the guidelines ($Mean = 2.67, SD = 0.83$), and (4) the attractiveness (from the general audiences) of the storyboards using the guidelines ($Mean = 4.57, SD = 0.74$) and the storyboards not using the guidelines ($Mean = 3.91, SD = 0.70$). For storyboards about the dataset \textit{The global trends in overweight and obesity}, the specific average scores include: (1) the creativeness of the storyboards using the guidelines ($Mean = 4.3, SD = 1.47$) and the storyboards not using the guidelines ($Mean = 3.8, SD = 2.87$), (2) the cinema of the storyboards using the guidelines ($Mean = 4.43, SD = 1.29$) and the storyboards not using the guidelines ($Mean = 2.97, SD = 0.86$), (3) the attractiveness (from experts) of the storyboards using the guidelines ($Mean = 4.6, SD = 0.87$) and the storyboards not using the guidelines ($Mean = 3.27, SD = 0.61$), and (4) the attractiveness (from the general audiences) of the storyboards using the guidelines ($Mean = 5.02, SD = 0.29$) and the storyboards not using the guidelines ($Mean = 4.51, SD = 0.40$).}
  \label{fig:comparative-score}
\end{figure*}

\subsection{Comparative Study Results}
This section shows the quantitative results of the scores in creativeness, cinematic expression (to what extent a storyboard could be regarded as being cinematic), and attractiveness (rated by both experts and the general audience) of the storyboards from the workshop (see ~\autoref{fig:comparative-score}) shows the average scores and 95\% confidence intervals for storyboards designed with and without our styles and guidelines. In both datasets, the scores of storyboards using our styles and guidelines are higher than those not using our styles and guidelines.

\subsubsection{Experts' Rating Results}
For each storyboard's creativeness, cinematic expression, and attractiveness score, we averaged the score from the three experts as its final scores. 
For each metric, we conducted the Mann-Whitney U test on the score of storyboards designed with and without our styles and guidelines as the normality assumption was violated.
\rv{Next, we report the $p-value$ and common language effect size (CLES) of the test results.}
Results show that the scores of creativeness ($p = 0.008, CLES = 0.825 $), cinematic expression ($p = 0.024, CLES = 0.765$), and attractiveness ($p = 0.013, CLES = 0.8$) of storyboards using our styles and guidelines were significantly higher than those of storyboards not using our styles and guidelines.
Given that each participant only designed the opening using one dataset, for the data related to each dataset, we also performed the Mann-Whitney U test on the three metrics. Results for each dataset are as follows: (1) \textit{Top 10 causes of death}: There was no statistically significant difference between the score of creativeness ($p = 0.104, CLES = 0.76$) and cinematic expression ($p = 0.145, CLES = 0.72$) of storyboards from \textit{guideline participants} and \textit{non-guideline participants}. However, the attractiveness of storyboards from \textit{guideline participants} was significantly rated higher than that of storyboards from \textit{non-guideline participants} ($p = 0.085, CLES = 0.78$) . (2) \textit{The global trends in overweight and obesity}: the storyboards from \textit{guideline participants} were rated significantly higher in terms of their creativeness ($p = 0.06, CLES = 1$), cinematic expression ($p = 0.018, CLES = 0.92$), and attractiveness ($p = 0.01, CLES = 0.96$) than the storyboards from \textit{non-guideline participants}.

\subsubsection{General Public's Rating Results}
Following the same statistical analysis procedure above, we found that generally, the attractiveness of storyboards from \textit{guideline participants} was significantly rated higher than storyboards from \textit{non-guideline participants} ($p = 0.041, CLES = 0.735$). Among the storyboards that used the dataset \textit{Top ten causes of death}, those that applied our styles and guidelines were not significantly rated as more or less attractive than storyboards that did not apply our styles and guidelines ($p = 0.148, CLES = 0.72$). However, among storyboards that used the dataset \textit{The global trends in overweight and obesity}, those that applied our styles and guidelines were significantly rated as more attractive ($ p = 0.037 CLES = 0.86$).

%% file: sections/7-discussion.tex
\section{Discussion}
\rv{The results of the workshop and the comparative study confirm that the openings designed with our guidelines were more attractive. In this section, we discuss the significant findings from our studies.}
\subsection{From 'Wow' to 'Why': Making Data Video Openings Engaging}
Our proposed styles and guidelines serve the purpose of enriching the styles of data videos that currently focus on explaining data insights ("Why") with new styles that attract the audience first ("Wow") and then explaining insights in detail.
Although the six cinematic opening styles function differently, all these styles essentially serve one common goal, which is to create a "Wow" effect in the opening to capture the audience's attention. The impact of the use of symbolism, motion graphics, shape, and color (\textit{Symbolism and Metaphor}), spectacular camera movement (\textit{Camera Eye}), visual effects of an explosion (\textit{Big Bang}), visual contrast between then and now (\textit{Old Footage}) with a preview of showing part of the ending first (\textit{Ending Firs}), shows that these styles attempt to create a "Wow" effect in their own different ways. \rv{While these six opening styles do not overlap with each other, they can be used in combination. However, combing them does not necessarily mean that the extent of the "Wow" effect will automatically increase proportionally to the number of styles applied.} 
As people's attention span becomes shorter nowadays in the environment full of distractions, capturing attention and evoking people's curiosity about data ("Wow") should be the first step for a data video before it proceeds to explain the insights ("Why"). 
With these opening styles in mind, we recommended that data videos should consider putting the creative or cinematic "Wow" first before the "Why".


\subsection{How Participants Choose Opening Styles}
The workshop results indicate that all the six opening styles were used by the \textit{guideline-participants}, and their frequencies were shown in Section~\ref{sec:workshop-result}. Both qualitative and quantitative feedback from \textit{guideline participants} indicated that they found each style effective in attracting the audience (the average score of the attractiveness of each style was higher than 5 on a 7-point scale). However, we found that some styles (e.g., \textit{Symbolism and Metaphor (6)}, \textit{Camera Eye (5)}) were used more often, while some (e.g., \textit{Big Bang (3)}, \textit{Ending First (1)}) were less frequently used by \textit{guideline participants}. When we asked \textit{guideline participants} how they selected the opening styles they used, they gave two main considerations. 
First, they selected styles that helped convey the topics or key messages of their stories. For example, two participants who used the dataset \textit{The global trends in overweight and obesity} intended to remind people of the importance of a healthy lifestyle. They both used the style \textit{Symbolism and Metaphor} by using people icons with a ``fat shape'' as the symbol of the story topic ``overweight and obesity''. They did so because they believed that human-shaped icons could emotionally connect to the audience and raise their awareness of physical health. 
Second, they first imagined what kinds of impact their openings should have and then selected suitable styles. For example, participants expected the effects of their data video openings differently, such as being suspenseful, surprising, or delightful.
One future work is to refine the framework of the styles and guidelines from a task-oriented perspective by mapping those items onto different needs of data video makers.

Interestingly, 7 out of 10 \textit{guideline participants} spontaneously and creatively combined different styles in their data video opening designs. Similarly, our corpus of data videos and films showed the combined uses of these styles. Although these opening styles could work together to consolidate the overall cinematic effects in an opening, combining more styles does not automatically lead to a more compelling opening in return. Future research could identify more effective styles and combinations by expanding our dataset and conducting larger-scale experiments.

We also analyzed the storyboards from \textit{non-guideline participants} \rv{and found that two out of 10 participants spontaneously used our styles and guidelines, whose works received at least a score of 4 or 5 from all the experts.}
\st{While experts' ratings on the storyboards from \textit{non-guideline participants} were mostly lower than 4, we found two storyboards received at} 
For example, one participant who used the dataset \textit{The global trends in overweight and obesity} showed an image of a devil before presenting the data about trends in overweight to symbolize the negative effects of being overweight. \rv{By contrast, other participants' works were mostly rated lower than 4. This result, from a different perspective, suggested the effectiveness of our guidelines.}
\st{It leaves an open question of whether styles that are more intuitive or popular would actually be more effective than others. Echoing our discussion above, a further question is whether the effectiveness of different styles varies according to the different needs of data video makers. Extensive evaluation studies are required in future work to answer these questions.}

\subsection{The Challenges of Applying the Styles and Guidelines}
Our participants mentioned two main challenges of applying our styles and guidelines. First, when achieving a style, they felt constrained by the literal meaning of the name of the style. For example, they assumed that \textit{Old Footage} only suggested that this opening style only uses newsreel or historical footage. Yet, film openings with flashback scenes other than newsreel footage showing the backstory of a story were also grouped in this category. After a reflection, we believe that a more complete, accurate name such as ``Old Footage/Backstory'' was more helpful for inspiring designers. For the issue of clarity, the name of this opening style remains unchanged throughout the paper. Another helpful approach could be providing more examples to enhance their inspiration. Future research in providing guidance on designing data stories could put more effort into how to inspire users' design ideas.
Second, although we had provided an application example of each guideline for participants, they mentioned that sometimes they had difficulty figuring out other ways to use a guideline. They also worried whether and how they could implement the desired visual effects through software. In our future work, we plan to provide extracted design templates from outstanding films and data videos to help data video makers select and realize each guideline. Moreover, we plan to conclude specific technical tutorials for realizing each guideline.

\subsection{Implications for Data Video Authoring Tools}
During our workshop, we asked participants what functions they would desire from future tools for assisting them in creating data stories. 
Most of them suggested tools that could recommend design templates with examples. 
Although one stream of research in narrative visualization tools studies how to provide users with design templates, their templates were only extracted from exiting data stories~\cite{amini2016authoring,sultanum2021leveraging,mcnutt2021integrated}. 
Our study results brought new design ideas from the film industry, which could be applied by future tools to generate new templates and examples. 
Moreover, one participant with a design background suggested that beyond tools, what designers wished for were platforms to share and learn ideas and skills. 
For example, many online communities provide learning resources such as tutorials for filmmaking software and stock footage for video and film producers. 
The narrative visualization community could build a platform referencing successful communities from the film or design industry. 
To sum up, we encourage future work in developing tools and platforms for data storytellers to put more effort into empowering them to learn skills and boost their inspiration.

%% file: sections/8-limitations-futurework.tex
\section{Limitations and Future Work}

\rv{\textbf{Styles and Guidelines.}} 
\rv{We believe the styles and guidelines are complete and comprehensive. However, we only studied a limited set of representative films and data videos. Future work should further evaluate the styles and guidelines against more films and data videos to check their completeness.}
\st{First,} While the \textit{guideline participants} in the workshop were satisfied with our proposed six styles, we believe our framework of styles and guidelines could be expanded by: (1) identifying more effective combinations of the styles for creating attractive data video openings, (2) connecting the styles and guidelines with the high-level needs of data video makers, such as conveying different emotions, and (3) adding easy-to-use design templates and technique tutorials to help designers apply each guideline.

\rv{\textbf{Experiment Designs.}} \st{Second,} In the experiment, to avoid overwhelming the participants with too much workload so that they could invest fair time and energy in designing video openings with and without our teaching material, we chose a between-subjects study design. However, such a design made it hard to assess whether and to what extent our guidelines help users improve their abilities to design attractive data video openings. For example, some \textit{non-guideline participants} created data video openings with our proposed styles on the basis of their past experiences.
Future work could conduct a within-subjects study that provides sufficient time between two trials to more rigorously evaluate the effectiveness of our teaching materials in helping users improve their abilities to design attractive data video openings.
\st{Third,} Additionally, we only did an in-lab study, in which participants used our guidelines in a limited amount of time. Little is known about how the guidelines could help data video makers in long-term practices. Future work should conduct long-term evaluation studies to uncover more potential challenges and issues that users may face in their daily work.

\rv{\textbf{Other Narrative Stages.}} \st{Fourth,}The opening is the first and one stage of the narrative of a data video. Future work should investigate ways to facilitate the design of other narrative stages, such as the climax~\cite{freytag1895technique} and the ending, which are also critical to the success of a data video. 

\rv{\textbf{Evaluation Metrics.} Our evaluation metrics focused on the primary goal of data video opening---attracting the audiences. However, we acknowledge that the ultimate goal of an entire data video is likely to communicate insights, convincing the audience, or having an affective impact. As a result, making data videos more memorable and understandable for the audience should be evaluated for the success of a data video in its entirety.}            

\rv{\textbf{Other Narrative Visualization Genres.} We observed that some of our styles and guidelines might also be applicable to the design of attractive openings for other narrative visualization genres that share similar properties with data videos. For example, both \textit{data comics} and \textit{data-GIFs} consist of a sequence of frames to convey ideas from data. Thus, some styles and guidelines that do not involve the use of sounds, such as the style Symbolism and Metaphor and guidelines G1.3 and G1.6, are also adaptable to data comics and data GIFs. Future work can systematically investigate ways to expand and evaluate our opening styles and guidelines to other narrative visualization genres.}

\rv{\textbf{Future Authoring Tools.}}\st{Finally,} Although we provide styles and guidelines, selecting and implementing them in data videos still requires extensive manual work from users. Therefore, future work should design and develop AI approaches~\cite{wu2021ai4vis} and (semi)-automatic authoring tools to recommend appropriate guidelines or provide templates for end-users to realize their designs.


%% file: sections/9-conclusion.tex
\section{Conclusion}
This work explores how to create attractive data video openings with cinematic styles.
We concluded six dominant cinematic opening styles that are adaptable to data videos from analyzing hundreds of films and data videos.
We further derived 28 guidelines for helping designers apply the six opening styles to data videos by consulting with experts from the film industry.
Our guidelines considered the design of both the narrative and visual representations of the opening. Besides the definition, each guideline was further illustrated by its core idea, execution tips, and visual examples to facilitate the learning process.
Our workshop and comparative study results showed that participants understood our styles and guidelines well, and the design of data video openings from participants who used our styles and guidelines were more compelling than those from participants who did not. 
We hope that our work can advance the body of knowledge on how to tell an attractive data story, help data video designers convey their data stories engagingly, and help future data video authoring tools develop design templates for users.